\documentclass[aps,prc,twocolumn,superscriptaddress,10pt,english,preprintnumbers]{revtex4-1}
\pdfoutput=1
\usepackage[T1]{fontenc}
\usepackage[utf8]{inputenc}
\usepackage{epsfig}
\usepackage{graphicx}
\usepackage{xcolor}
\usepackage{latexsym}
\usepackage{textcomp}
\usepackage{amssymb}
\usepackage[colorlinks=true,linkcolor=blue,citecolor=blue, urlcolor=blue]{hyperref} 
\usepackage{amsfonts,amsthm,amstext,amscd}
\usepackage{amsmath}
\usepackage{mathtools}

\DeclarePairedDelimiter\ceiling{\lceil}{\rceil}
\DeclarePairedDelimiter\floor{\lfloor}{\rfloor}
\def\trento{T\raisebox{-0.5ex}{R}ENTo}

\def\cent{cent_{\rm norm}}
\def\tr{\emph{Trajectum}}
\def\sf{\sigma_\text{fluct}}
\def\cs{\chi_\text{struct}}
\def\eos{(\eta/s)_\text{min}}
\def\mpt{\langle p_T \rangle}

\DeclareMathOperator{\sgn}{sgn}

\begin{document}
\title{Predictions and postdictions for relativistic lead and oxygen collisions with \emph{Trajectum}}
\author{Govert Nijs}
\affiliation{Center for Theoretical Physics, Massachusetts Institute of Technology, Cambridge, MA 02139, USA}
\author{Wilke van der Schee}
\affiliation{Theoretical Physics Department, CERN, CH-1211 Gen\`eve 23, Switzerland}
\begin{abstract}
We introduce a global analysis of relativistic heavy ion collisions using \emph{Trajectum} of a significantly higher precision and including a new option to vary the normalization of the centrality estimator. We use the posterior distribution of our parameters to generate a set of high statistics samples that allows us to make precise predictions including statistical and systematic uncertainties estimated from the model parameter distribution. The results are systematically compared with experiment whereby we also include many observables not included in the global analysis. This includes in particular (extremely) ultracentral anisotropic flow and mean transverse momentum, whereby we find satisfactory agreement with experiment where data is available. Lastly, we compute spectra and anisotropic flow for oxygen-oxygen collisions performed at RHIC and to be performed at the LHC and comment on how these collisions may inform us on properties of the Quark-Gluon-Plasma.
\end{abstract}

\preprint{CERN-TH-2021-160//MIT-CTP/5333}

\maketitle

{\hypersetup{hidelinks}
\tableofcontents
}

\section{Introduction}
The collisions of heavy ions at the Relativistic Heavy Ion Collider (RHIC) at Brookhaven and the Large Hadron Collider (LHC) at CERN have led to an accepted picture of a short pre-hydrodynamic phase followed by a relativistic fluid composed of quark-gluon plasma (QGP) with remarkably small shear viscosity and finally a cascade of interacting hadrons \cite{Heinz:2013th,Romatschke:2017ejr, Busza:2018rrf, Nagle:2018nvi, Berges:2020fwq}\@. An increasingly precise picture of these phases has recently emerged, including an accurate estimate of the temperature-dependent shear viscosity \cite{Bernhard:2016tnd,Moreland:2018gsh,Bernhard:2019bmu,Nijs:2020ors,Nijs:2020roc,JETSCAPE:2020shq,JETSCAPE:2020mzn}\@. Nevertheless, due to the complicated interplay of the various stages and corresponding models it has remained challenging to advance our quantitative understanding of heavy ion collisions.

There are multiple outstanding grand questions in the field of heavy ion collisions \cite{Busza:2018rrf}, which range from the limits of hydrodynamics to the behavior of energetic quarks and gluons as a function of energy scale and coupling constant. Crucial questions are the nature of the pre-hydrodynamic phase (both its lifetime and dynamics \cite{Berges:2020fwq}), the initial shape of the energy deposited by the colliding nuclei \cite{Snyder:2020rdy}, the behavior of hydrodynamics when viscous corrections are (too?) large \cite{Niemi:2014wta} or even if hydrodynamics can still make sense if only a few dozen particles per unit rapidity are produced. All these questions shed insights on a microscopic description based on QCD and can hence e.g.~inform how a strongly coupled QGP can arise from a theory of quarks and gluons which are free at asymptotically high energies.

Given our current state of understanding it is often difficult to shed light on these often quantitative questions by studying a single aspect and linking it to a single measurement. Instead, for instance both the viscosities and the initial shape characteristically alter the (anisotropic) momentum distribution of the final state particles. The initial shape and initial momentum anisotropies furthermore depend on the pre-hydrodynamic far-from-equilibrium stage. It is hence clear that only a global analysis taking into account all these questions and a representative set of experimental observables can put strong quantitative constraints on individual parameters such as the shear viscosity over entropy density ratio.

The aim of the current work is two-fold. Firstly, we would like to assess our current understanding of heavy ion collisions, including the systematic uncertainties that necessarily arise from modelling its different aspects. Modelling uncertainties are especially well addressed by studying the robustness of the model under widening of the scope of the model, such that the model has the flexibility to be closer to reality. This approach has recently been implemented by widening the initial shape \cite{Moreland:2018gsh}, the initial stage \cite{Nijs:2020ors}, allowing the variation of second order transport coefficients \cite{Nijs:2020ors, JETSCAPE:2020mzn}, widening the parameterization of the shear and bulk viscosities \cite{JETSCAPE:2020mzn} and studying several different viscous particlization prescriptions \cite{JETSCAPE:2020shq, Everett:2021ulz}\@. In this work we also add one parameter (the centrality normalization), but instead we focus on the existing model and perform a high statistics PbPb analysis of the model output at twenty randomly drawn model parameters from the posterior. This not only allows us to go beyond the experimental data used in the global analysis, but also allows an analysis of the systematic uncertainties due to our uncertainties in the model parameters. Discrepancies with the detailed experimental comparison then indicate a lacuna in our current understanding.

A second aim is to highlight smaller collision systems \cite{Nagle:2018nvi}\@. For this we do a similar analysis, but now for oxygen-oxygen (OO) collisions \cite{Brewer:2021kiv}. In May 2021 RHIC has performed a 16 day run with 402M OO collisions at a nucleon-nucleon collision energy of $\sqrt{s_{\rm NN}} = 0.2\,$TeV and LHC has a special run planned likely in 2024 at $\sqrt{s_{\rm NN}} = 6.8\,$TeV\@. No analyzed data is yet available and hence in contrast to the detailed PbPb analysis (mostly postdictions) the results of the OO analysis will include only predictions. Moreover, OO collisions are right at the threshold where a hydrodynamic QGP may well be present, but significant non-hydrodynamic effects are also expected. Predictions can hence inform us on which aspects are consistent with a hydrodynamic QGP\@.

Our analysis is performed using \tr{}, which is a new computational code \cite{Nijs:2020ors,Nijs:2020roc} to simulate heavy and light ion collisions that is entirely self-contained (with the exception of the hadronic gas phase) and includes one of the most versatile models for both the initial stage (based on the \trento{} model \cite{Bernhard:2016tnd}) and a hydrodynamic phase including temperature dependent first order transport coefficients and all second order coefficients. Importantly, it comes with an analysis routine that is fast and fully parallelizable, thereby allowing to analyze an unlimited number of events with methods that are designed to be as close to experimental measurements as possible.

In this work we present several improvements with regard to previous work. First of all the simulation of \tr{} in this work now has improved precision in its emulation of heavy ion collisions as a function of the parameter space whereby emulation uncertainties are now competitive or even significantly smaller than the corresponding experimental uncertainties. Secondly, we introduced a new parameter rescaling the anchor point (AP) for the 100\% centrality point. This has direct consequences for all non-central collisions and we show a significant change in the posterior of for instance the nucleon width. %

Lastly, as stated we present a multitude of predictions (or postdictions) for lead-lead (PbPb) collisions at the LHC as well as oxygen-oxygen (OO) collisions at both RHIC and the LHC\@. These predictions include a systematic uncertainty analysis from the full posterior distribution as well as statistical uncertainties for observables where this is relevant. Especially interesting are the (extremely) ultracentral results (down to 0.001\% centrality), where measurements are mostly not available yet. For the OO results we present estimates how the upcoming experimental results can help in improving our understanding of the QGP\@.

\section{Model}

For the heavy ion simulations presented in this work, we use version 1.1 of the \emph{Trajectum} code \cite{trajectumcode}\@.
As \emph{Trajectum} is described in some detail in \cite{Nijs:2020roc}, we will only give a brief outline here, highlighting the changes to the version of \emph{Trajectum} presented in \cite{Nijs:2020roc}\@.
\emph{Trajectum} allows the user to choose different models for the initial conditions, hydrodynamics and transport coefficients, and also offers a variety of PDE solvers for the hydrodynamical evolution.
As in \cite{Nijs:2020roc} we use the T\raisebox{-0.5ex}{R}ENTo model for the initial conditions, modified with a free streaming velocity that can differ from the speed of light.
For hydrodynamics we use the 14-moment approximation, with transport coefficients that allow for temperature dependence in the shear- and bulk viscosity over entropy density ratios, as in \cite{Bernhard:2019bmu}\@.
The hydrodynamical evolution is solved using the `fast' version of the MUSCL algorithm \cite{Nijs:2020roc}\@.

This leaves us with the following set of parameters that are varied in this work.
From T\raisebox{-0.5ex}{R}ENTo, we have the minimal inter-nucleon distance $d_\text{min}$, the number of constituents $n_c$, and the nucleon width $w$\@.
We also have the constituent width $v$, which is parameterized by the parameter $\chi_\text{struct}$, in terms of which $v$ is defined as follows:
\[
v = v_\text{min} + \chi_\text{struct}(w - v_\text{min}),
\]
where we fix $v_\text{min} = 0.2\,$fm.
The constituents in each colliding nucleon each then source a thickness function with a gaussian distribution, where for each individual constituent the norm of the gaussian is given by $N\gamma/n_c$\@.
Here $\gamma$ is sampled from a gamma distribution with width $\sigma_\text{fluct}\sqrt{n_c}$, and $N$ and $\sigma_\text{fluct}$ are parameters.
In this way, for each nucleus a thickness function is computed, $\mathcal{T}_A$ and $\mathcal{T}_B$, respectively\@.
The two thickness functions $\mathcal{T}_A$ and $\mathcal{T}_B$ so obtained are then combined to form the initial energy density $e$ according to
\begin{equation} \label{eq:trento}
e \tau = \left(\frac{\mathcal{T}_A^p + \mathcal{T}_B^p}{2}\right)^{1/p},
\end{equation}
with $\tau=0^+$ the initial proper time and and $p$ another parameter.

The pre-hydrodynamic stage performs free streaming with a velocity $v_\text{fs}$ until proper time $\tau_\text{fs}$, which are the two parameters governing this stage.
For the hydrodynamic stage, the temperature dependent shear viscosity over entropy density $\eta/s$ is described by the parameters $(\eta/s)_\text{min}$, $(\eta/s)_\text{slope}$ and $(\eta/s)_\text{crv}$, while the analogous parameters for the bulk viscosity $\zeta$ are $(\zeta/s)_\text{max}$, $(\zeta/s)_\text{width}$ and $(\zeta/s)_{T_0}$\@.
In addition, we vary the shear relaxation time $\tau_\pi$ by setting $\tau_\pi sT/\eta$ as a temperature independent parameter, and we vary the second order transport coefficient $\tau_{\pi\pi}$ by the parameter $\tau_{\pi\pi}/\tau_\pi$\@.
The final parameter that we vary is the freeze-out temperature $T_\text{switch}$, which enters in the particlization procedure (following \cite{Bernhard:2016tnd})\@.
In the following, we will describe the changes that we made to the model for this work.

\subsection{Oxygen nuclei}

To simulate a Pb nucleus, we use a Woods-Saxon distribution with parameters as given in \cite{Bernhard:2019bmu,DeVries:1987atn}, while enforcing a minimal distance $d_\text{min}$ between nucleons.
For a large spherical nucleus such a model is appropriate, but for smaller nuclei like O such an approach fails to capture important correlations between nucleon positions. Indeed, the 16 nucleons in oxygen tend to group together in four $\alpha$-particles, which potentially has important consequences for the shape of the geometry.
For this reason, for O we randomly rotate a sampled oxygen nucleus from a list of 6000 O configurations computed using effective two- and three-nucleon potentials, which is able to capture these correlations \cite{Lonardoni:2017egu,Alver:2008aq,Rybczynski:2019adt}\@. The difference between a standard Woods-Saxon distribution and the $\alpha$-clustered version has recently been studied in \cite{Rybczynski:2019adt, Summerfield:2021oex}\@.

\subsection{Continuous number of constituents}\label{sec:contnc}

In the T\raisebox{-0.5ex}{R}ENTo model with substructure \cite{Moreland:2014oya,Moreland:2018gsh}, the number of constituents $n_c$ is a discrete parameter.
In a Bayesian analysis, this can potentially cause problems for the emulator, since the emulator essentially interpolates between  different design points assuming that the quantities it emulates depend continuously on the parameters.
However, since $n_c$ is discrete, in the T\raisebox{-0.5ex}{R}ENTo model the dependence on $n_c$ is more like a stepwise defined function. Especially from $n_c = 1$ to $n_c = 2$ the effect on certain observables can be significant, which can cause the emulator to be less accurate.

In this work, we resolve this issue by changing the number of constituents to a continuous parameter.
Each sampled nucleon either uses $\floor{n_c}$ or $\ceiling{n_c}$ constituents with probability such that the average number of constituents equals $n_c$\@.
The definition agrees with the old definition for integer values of $n_c$, but for a large number of events this definition indeed leads to observables depending continuously on $n_c$\@.

\subsection{UrQMD and SMASH}\label{sec:URQMD}

Whereas \cite{Nijs:2020roc} used SMASH \cite{Weil:2016zrk,dmytro_oliinychenko_2020_3742965,Sjostrand:2007gs}, in this work we use UrQMD \cite{Bass:1998ca,Bleicher:1999xi} as a hadronic afterburner.
These codes are both state of the art, and are hence both well suited to be used in conjunction with \emph{Trajectum}\@.
Both codes produce similar results \cite{JETSCAPE:2020mzn}, whereby UrQMD leads to somewhat higher multiplicities (about 1\%) and higher $\langle p_T\rangle$ for kaons and protons (about 1.5\%)\@. There is no $\langle p_T\rangle$ increase for pions, but low-$p_T$ yields ($<0.4\,$GeV) are enhanced by about 1.5\%, as are high-$p_T$ ($>1.5\,$GeV) yields. The UrQMD elliptic flow is also slightly higher (1\%) across all centrality classes.

For the PbPb results presented in this work, it will turn out that the fitted free streaming velocity $v_\text{fs}$ is substantially smaller than the previous result obtained in \cite{Nijs:2020ors,Nijs:2020roc} (this can  compensate for higher $\langle p_T \rangle$)\@. The fitted minimum shear viscosity over entropy density ratio $(\eta/s)_\text{min}$ is slightly higher than the result obtained in \cite{Nijs:2020ors,Nijs:2020roc} (this can compensate for the larger elliptic flow)\@.
Even though the differences between UrQMD and SMASH seem mild they are indeed still the likely cause of the slightly changed posterior distributions, especially as the mean transverse momentum is a sensitive observable that is known experimentally to high precision (see also Fig.~\ref{fig:mult} later)\@.
This probably indicates that there is some work remaining to be done to understand the correct way to describe the final stages of the collision, but at least between UrQMD and SMASH the discrepancies are quite mild. 

\subsection{Centrality determination}

Historically the determination of centrality classes has not always been clear. Theoretically often the impact parameter of the collision was used as a probe for centrality, with smaller impact parameter giving more central collisions. Fluctuations in geometry and multiplicity give this method a significant bias and it is therefore much better to order centrality by the final multiplicity (ordering by initial entropy also gives reasonable results \cite{Schenke:2020mbo})\@. \tr{} has the option to specify several centrality definitions with different pseudorapidity $\eta$ and transverse momentum $p_T$ cuts to make these as close as possible to the actual experimental method. In this work we always work with the standard settings $|\eta| < 2.4$ and $p_T > 400\,$MeV$/c$ unless otherwise noted.

Experimentally the centrality ordering is usually defined by the number of tracks at forward rapidity (for ALICE the V0A and V0B detectors located at pseudo-rapidity $2.8 < \eta <5.1$ and $-3.7 <\eta < -1.7$ respectively)\@. This can lead to a problem comparing to theoretical models, which often work under the assumption of boost invariance and hence only compute multiplicities at mid-rapidity. Nevertheless, especially for PbPb collisions multiplicities at mid-rapidity are rather strongly correlated with the V0 amplitudes (see Fig.~15 in \cite{ALICE:2013hur}) and unless one looks at rather extreme centralities the methods should match fairly well. An example of this is explored in Fig.~\ref{fig:ultracentral}\@. For $p$Pb collisions this is however completely different and self-correlations in the centrality determination can be important \cite{ALICE:2014xsp}\@.

A single crucial question remains though: when do we determine that a collision happened? Experimentally this is far from easy: first of all the PbPb cross section is completely dominated by electromagnetic interactions of both charged nuclei. These interactions lead to the dissociation of a single ($187 \pm 12\,$b) or both ($5.7 \pm 0.4\,$b) Pb nuclei \cite{ALICE:2012aa}, which is much larger than the estimated hadronic cross section of $7.7 \pm 0.6\,$b. It is also necessary to correct for collisions that are hadronic but do not leave any tracks in the detector acceptance. Alternatively it is possible to fit the multiplicity distribution with a MC Glauber model and take the 100\% anchor point as given by the model, but this introduces a theoretical modelling uncertainty. Nevertheless, both methods agree well, and the final systematic uncertainty is less than 3\% \cite{ALICE:2013hur}\@.  

Also theoretically it is not quite clear when a hadronic collision actually happened. This is usually determined by a Monte Carlo Glauber model that puts nucleons at fluctuating positions and then decides in a collision whether or not two nucleons collide or not \cite{Miller:2007ri, dEnterria:2020dwq}\@. A crucial ingredient herein is the nucleon-nucleon cross section, $\sigma_{\rm NN}$, which also features as a \tr{} parameter and is usually fixed by the measured value in $pp$ collisions (at high energies the difference in cross section between protons and neutrons is negligible)\@. Nevertheless, several options are possible to satisfy the $\sigma_\text{NN}$ cross section, and a black disk approximation (as in MC Glauber \cite{Alver:2008aq,Loizides:2014vua}) does not lead to the same result as the \trento{} model with nucleons of size $w$ used by \tr{}\@.

In \tr{} (as in \trento{}) it is possible that a nucleon-nucleon interaction happened (i.e.~the number of participants is nonzero), but that the switching temperature is not reached anywhere and no thermal particles are produced (since the freeze-out is a Poisson process this may even happen if $T_\text{switch}$ is reached)\@. Nevertheless, since a hadronic interaction happened particles realistically should be produced  in a more complete model that takes the QGP corona into account and hence \tr{} does include these events in its event list as having zero particles.

While both the theoretical and experimental choices and procedures are well motivated and studied still some significant uncertainty remains and also to test the robustness of the model we decided to introduce a new parameter $cent_\text{norm}$ that rescales the centrality classes. This parameter rescales the $\alpha$--$\beta$\% centrality so that in the model we actually compute the $0.01\alpha \cdot cent_\text{norm}$--$0.01\beta \cdot cent_\text{norm}$\% centrality bin. In other words, we perform the computation as if 100\% experimental centrality corresponds to $cent_\text{norm}$ centrality in our model.

By including $cent_\text{norm}$ as a parameter, the uncertainty coming from this issue can be properly taken into account in the Bayesian analysis. That way, we can see which parameters correlate with $cent_\text{norm}$, and we are also able to fit $cent_\text{norm}$\@. It is however an interesting question whether in a way we may be double counting uncertainties here, since the experimental measurements already contain a contribution in their uncertainty due to the centrality selection. Since indeed we will find $cent_\text{norm}$ to be consistent with 100\% (see Section \ref{sec:posterior}) this could motivate to reinstate $cent_\text{norm} = 100\%$ in future analyses.

Important for the discussion in Sec.~\ref{sec:URQMD} is that $cent_\text{norm}$ is a parameter which enters purely in collecting the results of the (parallized) analysis of all events, not the simulation or analysis.
This means that it can easily be turned off at little extra computational cost, so we have been able to verify that the difference between this work and that obtained in \cite{Nijs:2020roc} is not due to the inclusion of $cent_\text{norm}$ (see also Fig.~\ref{fig:chainspbpb} later)\@.

\begin{figure*}[ht]
\includegraphics[width=0.99\textwidth]{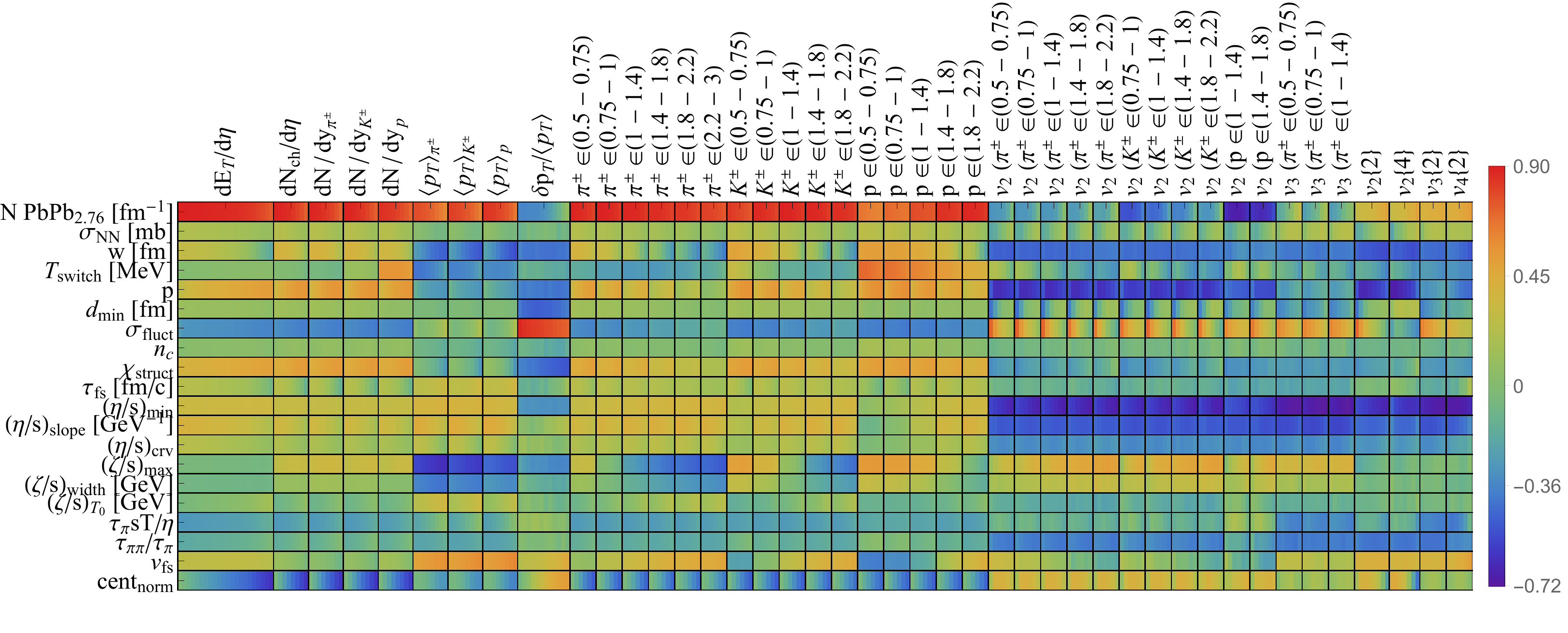}
\includegraphics[width=0.71\textwidth]{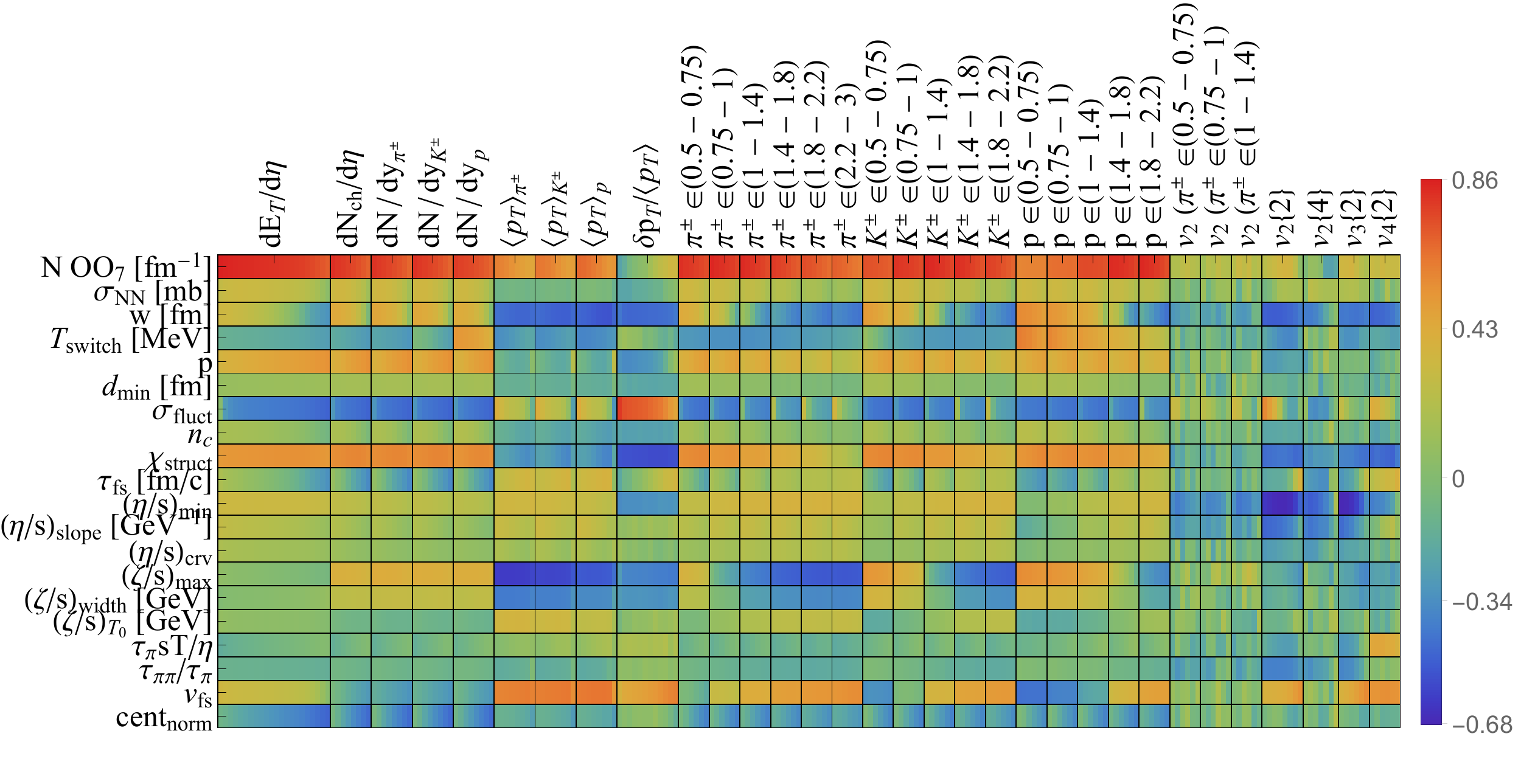}
\includegraphics[width=0.24\textwidth]{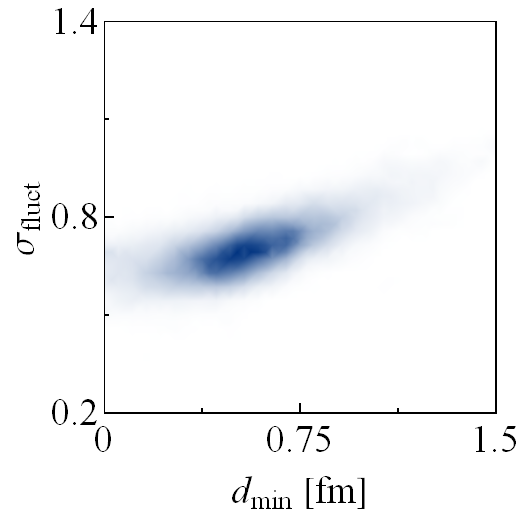}
\caption{\label{fig:design}We show the correlation between our parameters and all observables used for PbPb (top) and OO (bottom left) collisions. Inside each box different centralities are shown from most central on the left to most peripheral on the right. $p_T$-bins are displayed in units of GeV$/c$\@. (bottom right) The figure shows the posterior correlation between $d_{\rm min}$ and $\sf{}$ following from the posterior analysis done in Sec.~\ref{sec:posterior}\@.}
\end{figure*}

\section{Precise PbPb results and OO predictions}

\subsection{The design and emulator validation}\label{sec:design}

In this work, we perform a Bayesian analysis of the PbPb system and subsequently make (predominantly) postdictions for PbPb and predictions for the OO system. In the end we will also examine how much the precision of the model parameters can improve by the inclusion of future OO data. To make this as worthwhile as possible, we want to make the Bayesian analysis of both the PbPb and OO systems as precise as we can.  

The starting point of any Bayesian analysis is Bayes posterior probability density for our parameters
\begin{equation}
    \mathcal{P}(\boldsymbol{x}|\mathbf{y}_{\exp})
    = \frac{e^{-\Delta^2/2}}{\sqrt{(2\pi)^{n} \det\left(\Sigma(\boldsymbol{x})\right)}} \mathcal{P}(\boldsymbol{x}) 
    \label{eq:bayes}
\end{equation}
with $\mathcal{P}(\boldsymbol{x})$ the (flat) prior probability density and where
\begin{equation}
    \Delta^2
    = \left(\mathbf{y}(\boldsymbol{x})-\mathbf{y}_{\rm exp}\right)\cdot \Sigma(\boldsymbol{x})^{-1} \cdot \left(\mathbf{y}(\boldsymbol{x})-\mathbf{y}_{\rm exp}\right),
    \label{eq:delta}
\end{equation}
with $\mathbf{y}(\boldsymbol{x})$ the predicted data for parameters $\boldsymbol{x}$, $\mathbf{y}_{\rm exp}$ the $n$ experimental data points and $\Sigma(\mathbf{x})$ is the sum of the experimental and theoretical covariance matrices. In principle the full experimental matrix should be provided by the relevant experiment, but as this is rarely available we follow the simple prescription from \cite{Bernhard:2018hnz}\@. Obtaining the posterior distributions for our parameters involves an MCMC \cite{ForemanMackey:2012ig} that evaluates the model many times by use of a Gaussian Process emulator \cite{williams2006gaussian,Bernhard:2018hnz,Moreland:2019szz}\@. The uncertainty of the emulator is estimated by the emulator itself and included in $\Sigma$\@. For this analysis we use emulators only for the 50 most significant Principal Components (PCs) of the experimental data.

A crucial element in the emulator of a particular PC is the characteristic length scale it assigns to a parameter: the emulator uses shorter length scales when the PC is more sensitive to the parameter. This means it has to rely on fewer design points and hence typically has a larger uncertainty. By restricting the prior ranges of our parameters it is possible to make PCs less sensitive (on the domain) and hence a more precise emulator is expected. The price to pay is of course the smaller domain of the prior range, but from the previous work \cite{Nijs:2020roc} it was possible to tighten the prior range significantly without affecting our conclusions. Compared to \cite{Nijs:2020roc}, we furthermore lowered the statistical uncertainty by increasing the number of events per design point from 6k to 15k\@. Given the increase in computation time needed for this number of events per design point, we lower the number of design points from 1000 to 750 so that the total computation time remains manageable. Also, as discussed in Sec.~\ref{sec:contnc}, by making the number of constituents a continuous parameter, we no longer present the emulator with discontinuities. For OO collisions we ran the model at 1500 design points with 40k events each \footnote{The total computational time for the analysis was around 1M CPU hours, roughly equally divided over PbPb and OO collisions. For PbPb collisions half of the time was spent on UrQMD (the hadronic afterburner, we use an $\eta$ cut of 2.5 for this analysis), whereas for smaller OO systems almost all time was used by \tr{}\@.}\@.

\begin{figure*}[ht]
\includegraphics[width=0.75\textwidth]{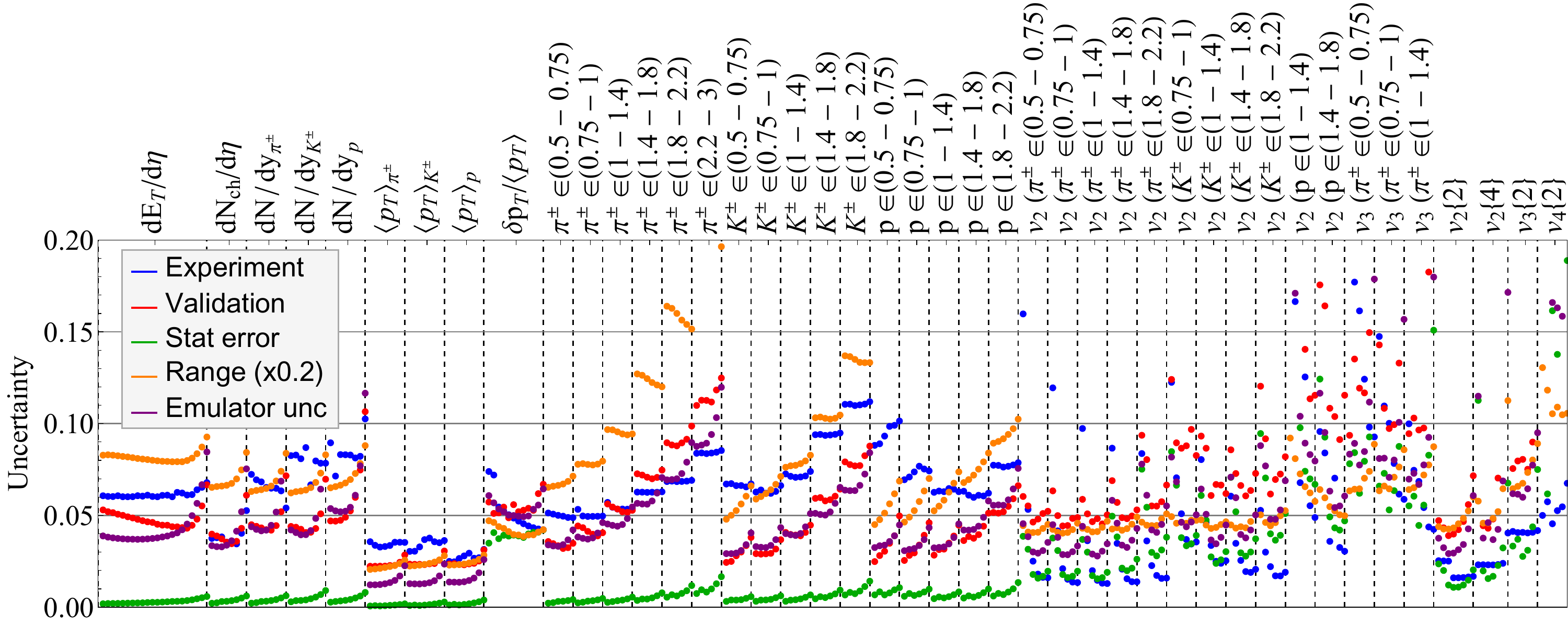}
\caption{\label{fig:emulatorerror} We show the validation of the emulator as a fractional uncertainty. The experimental, emulator and statistical errors are shown, as well as the prior range and the uncertainty reported by the emulator itself.}
\end{figure*}

One of the first interesting outputs to look for in the result of all 750 design points is the correlation between the various parameters and the observables, which is shown in Fig.~\ref{fig:design} for both PbPb (top) and OO (bottom)\@.
It can clearly be seen that total transverse energy, mean transverse momenta and (identified) ($p_T$-differential) particle yields correlate positively with the norm.
This makes sense, as the norm directly controls how much energy is deposited into the initial state.
As is well known the shear viscosity reduces all anisotropic flow coefficients $v_n\{k\}$ and the bulk viscosity suppresses the mean transverse momentum. As expected a higher switching temperature increases proton yields.
Perhaps unsurprisingly, the per-nucleon entropy deposition fluctuations $\sigma_\text{fluct}$ correlate strongly with fluctuations of mean transverse momentum $\delta p_T/\langle p_T\rangle$\@. Less obvious is that $\sigma_\text{fluct}$ also correlates positively with the most central bin of the anisotropic flow coefficients, to which we will come back in Sec.~\ref{sec:v2v3puzzle}\@.
The free streaming velocity $v_\text{fs}$ is also correlated with the mean transverse momentum, allowing both $(\zeta/s)_\text{max}$ and $v_\text{fs}$ to adjust to fit this observable.
However, whereas in PbPb $v_\text{fs}$ does not have a correlation with $p_T$-differential anisotropic flow, $(\zeta/s)_\text{max}$ is moderately correlated, explaining why the inclusion of $p_T$-differential observables in \cite{Nijs:2020roc,Nijs:2020ors} led to a substantially lower value for $(\zeta/s)_\text{max}$ than earlier work \cite{Bernhard:2016tnd,Bernhard:2019bmu,Ryu:2015vwa}\@.
The new parameter $cent_{\text norm}$ is clearly more important for peripheral collisions, especially for the yields. Note however, that for especially $v_2$ the centrality normalization is (relatively) important across the entire centrality range, which is explained by the observation that $v_2$ becomes very small towards central collisions.

A few parameters only have strong correlations with very few observables. A good example is $d_{\rm min}$ which strongly influences the transverse momentum fluctuations. This is however more heavily influenced by $\sigma_{\rm fluct}$, so that we will indeed only find very mild constraints on $d_{\rm min}$, albeit stronger constraints on the correlation between $d_{\rm min}$ and  $\sigma_{\rm fluct}$ (see Fig.~\ref{fig:design})\@. Lastly, a few parameters have only mild correlations, such as $\tau_\pi$ and $n_c$\@. This will mean that likely high precision will be needed to get constraints on those parameters. Note, however, that these figures can still hide important correlations when e.g.~two parameters interplay with each other.

The correlations for the OO figure show similar characteristics as the PbPb result, albeit likely showing higher statistical imprecision for the $p_T$-differential $v_2$ results. It is however clear that OO collisions due to their smaller size are more sensitive to the initial stage parameters such as $v_\text{fs}$ and $\tau_\text{fs}$\@.

A lower bound for the achievable precision in a Bayesian analysis is given by the experimental uncertainties on the data used in the analysis. Often, however, the theoretical emulation uncertainty is considerably higher.
In Fig.~\ref{fig:emulatorerror}, we show the validation of the emulator in this work, which one can compare to Fig.~4 of \cite{Nijs:2020roc}\@. One can see that both the statistical and emulator uncertainty have improved in this work, and for many observables the emulator uncertainty (which incorporates the statistical uncertainty as well) is now smaller than the experimental uncertainty, or comparable in size. This indicates that we are close to the most precise results which are achievable given current experimental uncertainties and the set of observables that we fit to. One potential issue is that the emulator's self-reported uncertainty (indicated in purple) tends to be lower than the error computed by comparing the emulator prediction to model computations (indicated in red)\@. This indicates that the emulator may be slightly underestimating the true theoretical error. Compared to \cite{Nijs:2020roc} the ranges of the observables are also decreased, which reflects the smaller prior ranges on our parameters. Nevertheless, given the previous analysis we were able to do so in a manner that still includes good coverage over the experimental data.

The experimental dataset that we fit to is the same as in \cite{Nijs:2020ors}, except that in this work we only fit to PbPb observables. In particular, we include charged particle multiplicity $dN_\text{ch}/d\eta$ at 2.76 \cite{Aamodt:2010cz} and 5.02 TeV \cite{Adam:2015ptt} and transverse energy $dE_T/d\eta$ at 2.76 TeV \cite{Adam:2016thv}\@. We also include identified yields $dN/dy$ and mean $p_T$ for pions, kaons and protons at 2.76 TeV \cite{ALICE:2013mez}, as well as $p_T$ fluctuations at 2.76 TeV \cite{Abelev:2014ckr}\@. For integrated anisotropic flow, we include $v_2\{2\}$, $v_2\{4\}$, $v_3\{2\}$ and $v_4\{2\}$ at both 2.76 and 5.02 TeV \cite{Adam:2016izf}\@. We also include $p_T$-differential observables, in six coarse grained $p_T$-bins with bin boundaries at $(0.5, 0.75, 1.0, 1.4, 1.8, 2.2, 3.0)\,\text{GeV}$\@. With these $p_T$-bins, we include identified transverse momentum spectra for pions, kaons and protons at 2.76 TeV \cite{Abelev:2013vea}, as well as the $p_T$-differential identified anisotropic flow coefficient $v_2\{2\}(p_T)$ for pions, kaons and protons at 2.76 TeV, and $v_3\{2\}(p_T)$ for pions \cite{Adam:2016nfo}\@. For the $p_T$-differential anisotropic flow, we only use bins with good statistical uncertainties, which are the same as the ones used in \cite{Nijs:2020ors}\@.

\begin{figure*}[ht]
\includegraphics[width=0.99\textwidth]{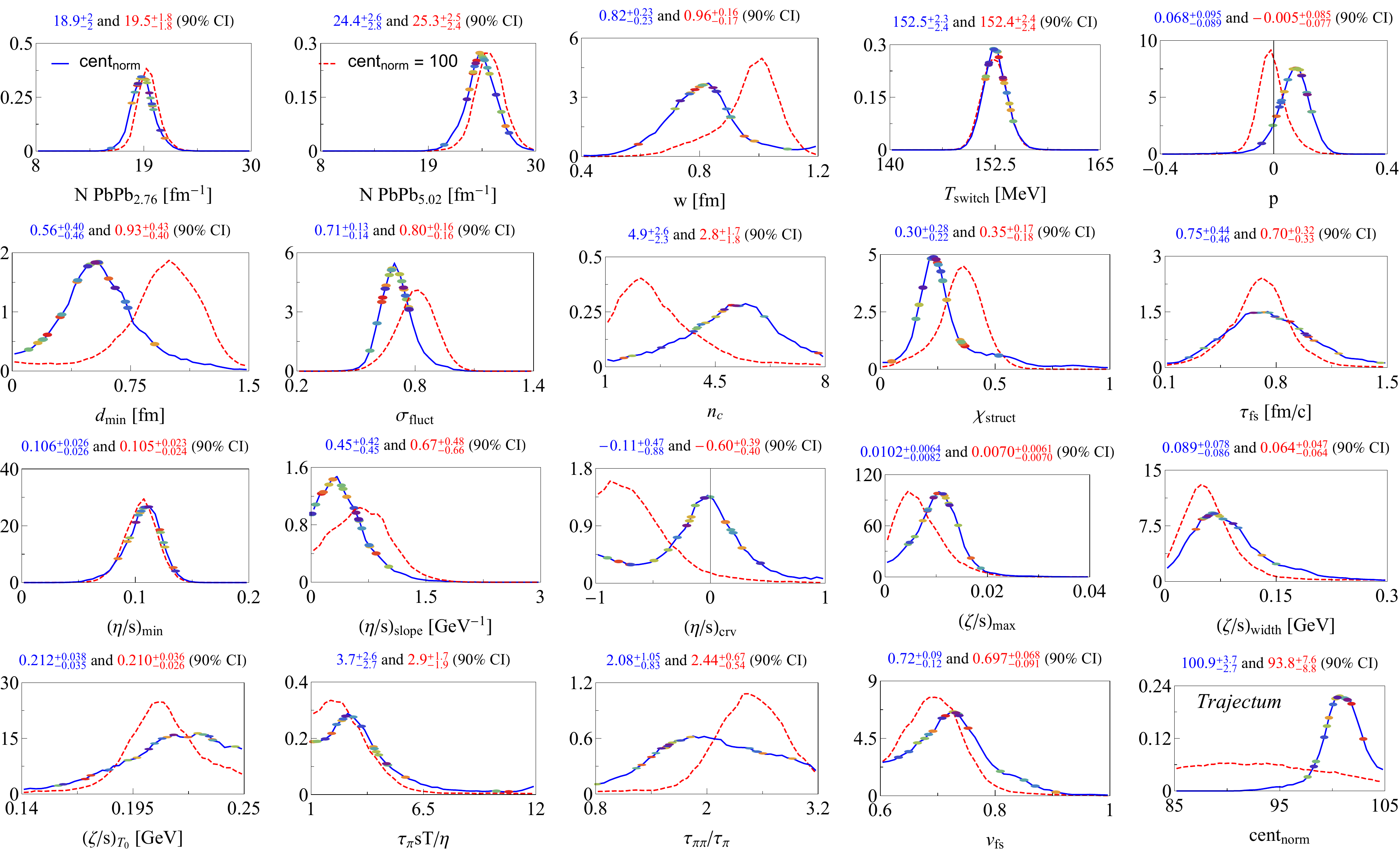}
\caption{\label{fig:chainspbpb}We show MCMC posterior distributions for our parameters fitted to PbPb collisions at collision energies of 2.76 and 5.02 TeV both with (solid blue) and without (dashed red) a varying centrality normalization $\cent$\@. From the distribution including variations in $\cent$ we randomly selected 20 sample points for which postdictions are shown in Sec.~\ref{sec:posterior}\@.
}
\end{figure*}

\subsection{Posterior distribution}\label{sec:posterior}

In Fig.~\ref{fig:chainspbpb}, we show the posterior distributions for the parameters from \eqref{eq:bayes} evaluated using the parallel tempered \cite{B509983H} emcee code \cite{ForemanMackey:2012ig} with three temperatures and 300 walkers applied to experimental data from PbPb collisions at 2.76 and 5.02 TeV\@. In particular, two different posteriors are shown. The solid blue lines correspond to a posterior in which $cent_\text{norm}$ was varied as a parameter, whereas in the dashed red lines $cent_\text{norm}$ is fixed to 100\% at the level of the \tr{} analysis\@. It can immediately be seen that the preferred value of $cent_\text{norm}$ is close to 100\%, with a relatively small uncertainty. This indicates that, within the posterior uncertainties, 100\% centrality in our model corresponds well with the experimental notion of 100\% centrality. Note however, that the posterior distributions for some observables, such as $w$ or $(\eta/s)_\text{crv}$, change substantially if $cent_\text{norm}$ is varied, indicating that small discrepancies between the model and experimental centrality determination can have a substantial effect on the posteriors of those parameters.

It is important to note that the red dashed line of $cent_\text{norm}$ is not completely flat, even though none of the observables depend on $cent_\text{norm}$ (it was fixed to 100\%)\@. Nevertheless, the emulator still associates the 750 design points to a given $cent_\text{norm}$ and attempts to find correlations between $cent_\text{norm}$ and the observables. Of course we could have explicitly told the MCMC analysis that our emulator does not depend on $cent_\text{norm}$ and that it hence could use a flat prior and posterior. In this work we chose not to do this to illustrate this effect and also to verify that indeed the MCMC does not exclude any of the $cent_\text{norm}$ prior distribution as should be expected. Also, for a flat prior and an ideal emulator we would expect to find the 90\% confidence interval to be given by $95\pm 9$, which is indeed very close to the result of $93.8^{+7.6}_{-8.8}$\@.

\begin{table}[ht]
\begin{tabular}{lrlr}
\hline
\hline
parameter & $I$ [\%] & parameter & $I$ [\%] \\
\hline
$N\text{ PbPb}_\text{2.76}$ & $50$ & $(\eta/s)_\text{min}$ & $50$ \\
$N\text{ PbPb}_\text{5.02}$ & $56$ & $(\eta/s)_\text{slope}$ & $97$ \\
$w$ & $-22$ & $(\eta/s)_\text{crv}$ & $21$ \\
$T_\text{switch}$ & $66$ & $(\zeta/s)_\text{max}$ & $-19$ \\
$p$ & $14$ & $(\zeta/s)_\text{width}$ & $40$ \\
$d_\text{min}$ & $22$ & $(\zeta/s)_{T_0}$ & $18$ \\
$\sigma_\text{fluct}$ & $44$ & $\tau_\pi sT/\eta$ & $21$ \\
$n_c$ & $0$ & $\tau_{\pi\pi}/\tau_\pi$ & $-15$ \\
$\chi_\text{struct}$ & $22$ & $v_\text{fs}$ & $33$ \\
$\tau_\text{fs}$ & $-17$ &  &  \\
\hline
\hline
\end{tabular}
\caption{\label{tab:pbpbimprovement}Improvement in posterior uncertainty comparing the posterior including a varying $cent_\text{norm}$ from this work (solid blue in Fig.~\ref{fig:chainspbpb}) to the combined PbPb and $p$Pb posterior from \cite{Nijs:2020ors}\@.}
\end{table}

Also, compared with Fig.~1 from \cite{Nijs:2020ors}, the posteriors of many of the parameters have become more constrained, due to the decreased statistical and emulator errors.
This can be seen in some detail in Tab.~\ref{tab:pbpbimprovement}, where we compare the posterior including a varying $cent_\text{norm}$ from this work to the combined PbPb and $p$Pb posterior from \cite{Nijs:2020ors}\@.
Here we define the improvement $I$ between the two works as
\[
I = \frac{\sigma_\text{\cite{Nijs:2020ors}}}{\sigma_\text{this work}} - 1,
\]
with $\sigma$ the 90\% confidence interval from each respective work.
One can see that the precision in the determination of both norms $N\text{ PbPb}_\text{2.76}$ and $N\text{ PbPb}_\text{5.02}$ improves substantially in this work, as well as the the switching temperature $T_\text{switch}$ and the three parameters relating to the shear viscosity over entropy density ratio $\eta/s$\@.
Of particular note is the parameter controlling the fluctuations in initial entropy deposition $\sigma_\text{fluct}$, which also has greatly improved precision, and which we will come back to in Sec.~\ref{sec:v2v3puzzle}\@.
Even though the improvement for $\tau_\pi sT/\eta$ is modest, it is worth noting that with the present posteriors, the value obtained from holography ($4 - \log(4) \approx 2.61$, \cite{Baier:2007ix}) is now clearly favored over the value obtained by a weakly coupled computation (5, \cite{Denicol:2014vaa}) for fixed $cent_\text{norm}$, and slightly favored with varying $cent_\text{norm}$\@.
As noted before, the posteriors for $v_\text{fs}$ and $(\eta/s)_\text{min}$ are substantially different from those obtained in \cite{Nijs:2020ors}, which is likely due to the difference between using UrQMD in this work, and using SMASH in \cite{Nijs:2020ors}\@.

\begin{figure}[ht]
\includegraphics[width=0.49\textwidth]{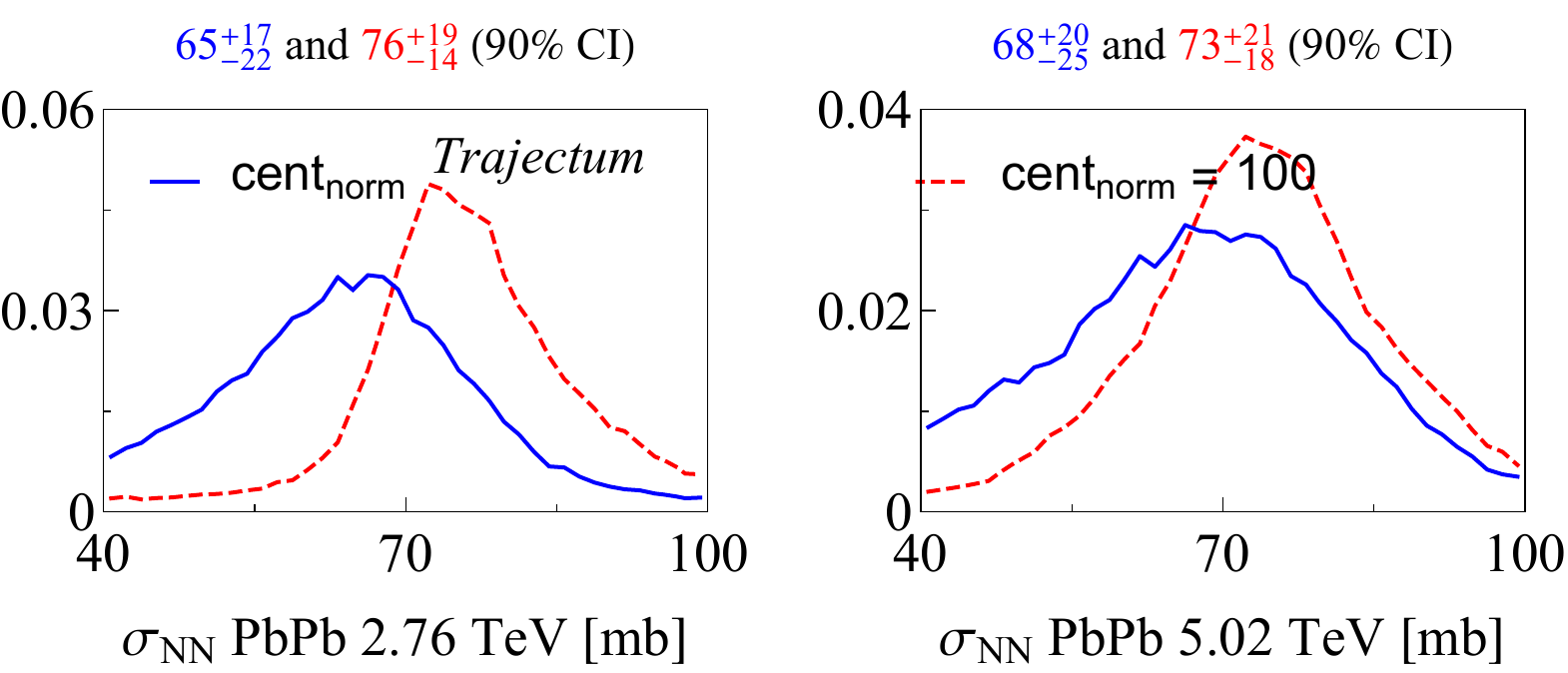}
\caption{\label{fig:sigmann}We show MCMC posterior distributions for the nucleon-nucleon cross section for both 2.76 (left) and $5.02\,$TeV (right) collision energies both with (solid blue) and without (dashed red) a varying centrality normalization $cent_\text{norm}$\@. All are consistent with the measured values of $61.8\pm 0.9$ and $67.6\pm 0.6\,$mb for 2.76 and $5.02\,$TeV respectively \cite{Loizides:2017ack}\@.
}
\end{figure}

In addition to the parameters shown in Fig.~\ref{fig:chainspbpb}, in the model design we also vary the nucleon-nucleon cross-section $\sigma_\text{NN}$\@. The reason for this is mainly that this does away with the need to perform separate runs for both collision energies \cite{Nijs:2020roc}\@. For a normal MCMC analysis we fix $\sigma_\text{NN}$ to 63 or 70 mb (for 2.76 and 5.02 TeV), but it is also possible to attempt to fit $\sigma_\text{NN}$ to the available data. This is shown in Fig.~\ref{fig:sigmann}, where the analysis shown in Fig.~\ref{fig:chainspbpb} was repeated while allowing $\sigma_\text{NN}$ to vary. Two interesting observations can be made. Firstly, the previously known values of $63\,\text{mb}$ ($\sqrt{s_\text{NN}} = 2.76\,\text{TeV}$) and $70\,\text{mb}$ ($\sqrt{s_\text{NN}} = 5.02\,\text{TeV}$) are compatible with our fits, both with and without varying the centrality normalization $cent_\text{norm}$\@. Secondly, the fit is not very sensitive to the value of $\sigma_\text{NN}$ at either collision energy. This indicates that one could probably produce reasonable predictions for both collision energies by fixing $\sigma_\text{NN}$ to a reasonable value compatible with the fits for both collision energies. It is interesting to see whether these fits can be made more precise in the future by including extra experimental data, such as the total hadronic PbPb cross section.

\subsection{Postdictions with systematic uncertainties}\label{sec:postdictions}

In addition to showing posteriors for the parameters, we can also use the posterior distributions to compute observables that we did not use in the fit. To do this, we drew 20 random sets of parameters (indicated in Fig.~\ref{fig:chainspbpb}) from the posterior, and computed our observables for all of them using 0.5M events per parameter set for PbPb collisions and 1M (2M) events for OO collisions at $\sqrt{s_\text{NN}}$ of 7 (0.2) TeV\@. For 7 TeV collisions we extrapolated the norm $N$ from the 2.76 and 5.02 norms by assuming a power-law scaling, whereas for 0.2 TeV we multiplied the 2.76 norm by 0.497 (this factor approximates the AuAu multiplicity at 0.2 TeV relatively well)\@. We used 39.7 and 74.2 for $\sigma_{\rm NN}$ for respectively 0.2 and 7 TeV collisions. These 20 sets allow to estimate a (one standard deviation) systematic uncertainty coming from our uncertainties in the model parameters in addition to the statistical uncertainty that is computed for all observables. Since we can use considerably more events as compared to the design points we can also make predictions or postdictions for more statistically demanding observables. %
Note that since we are drawing the parameters from the true posterior, and not just the 1D projections shown in Fig.~\ref{fig:chainspbpb}, all of the correlations between the posteriors of the parameters are automatically taken into account in the prediction.

\begin{figure*}[ht]
\includegraphics[width=0.34\textwidth]{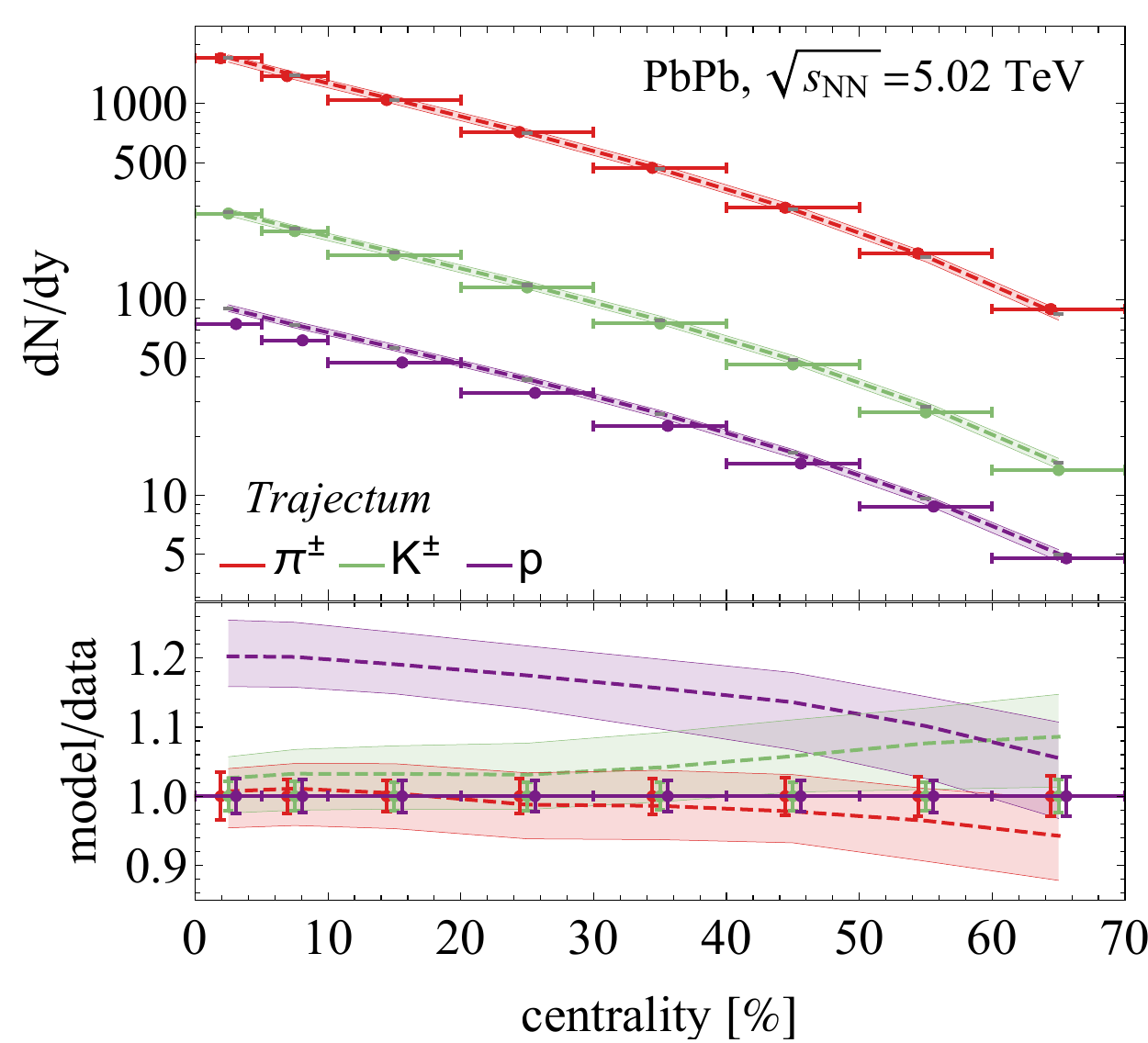}
\includegraphics[width=0.34\textwidth]{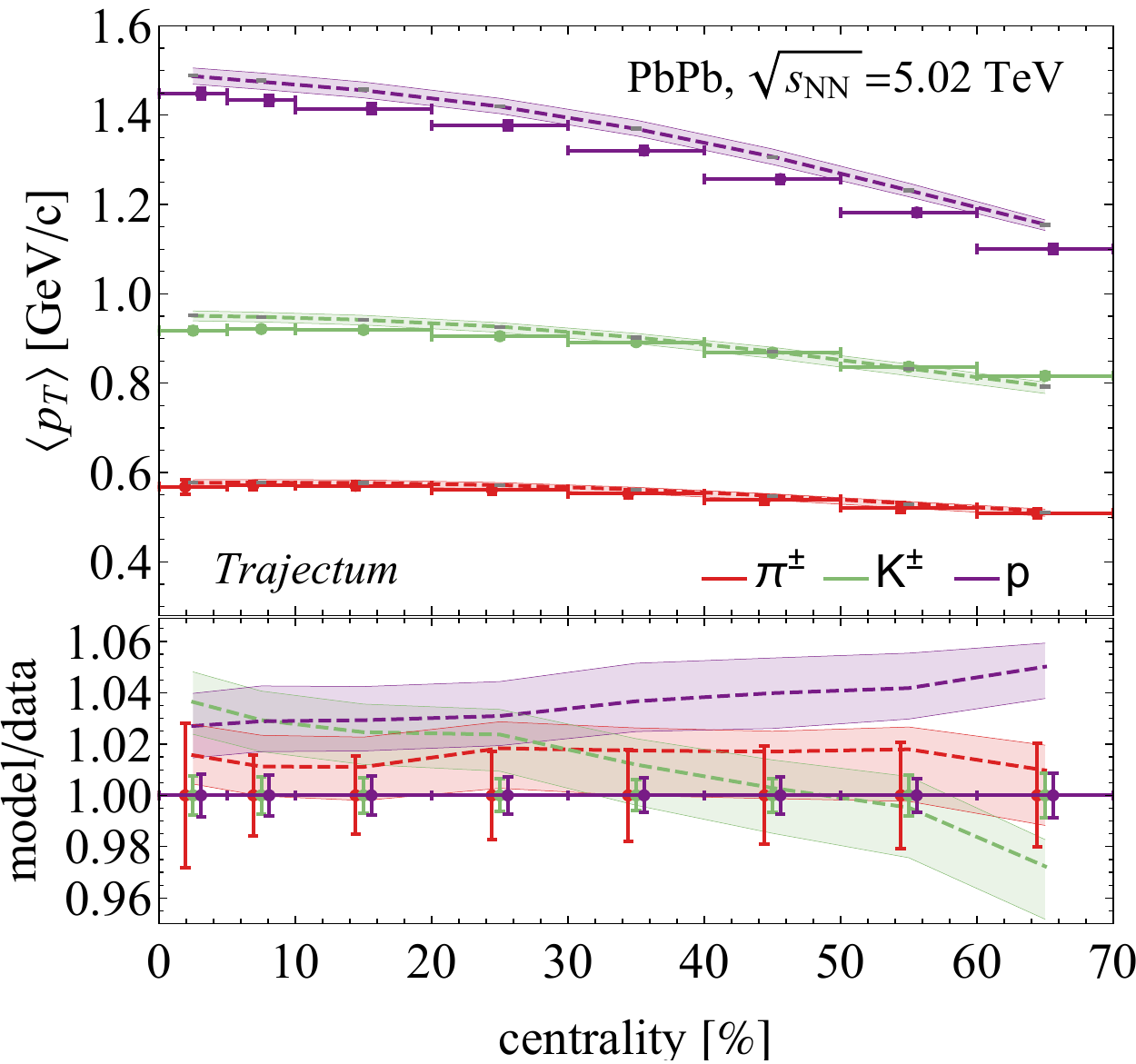}
\caption{\label{fig:mult}We show identified particle yields $dN/dy$ (left) and mean $p_T$ (right) for pions, kaons and protons at mid-rapidity as a function of centrality for PbPb collisions at $\sqrt{s_\text{NN}} = 5.02\,\text{TeV}$ compared to colored ALICE data points \cite{ALICE:2019hno}\@.
The bands represent systematic uncertainties (1$\sigma$) in the \emph{Trajectum} parameters (see Fig.~\ref{fig:chainspbpb}) and the grey points show statistical uncertainties.}
\end{figure*}

\begin{figure*}[ht]
\includegraphics[width=0.32\textwidth]{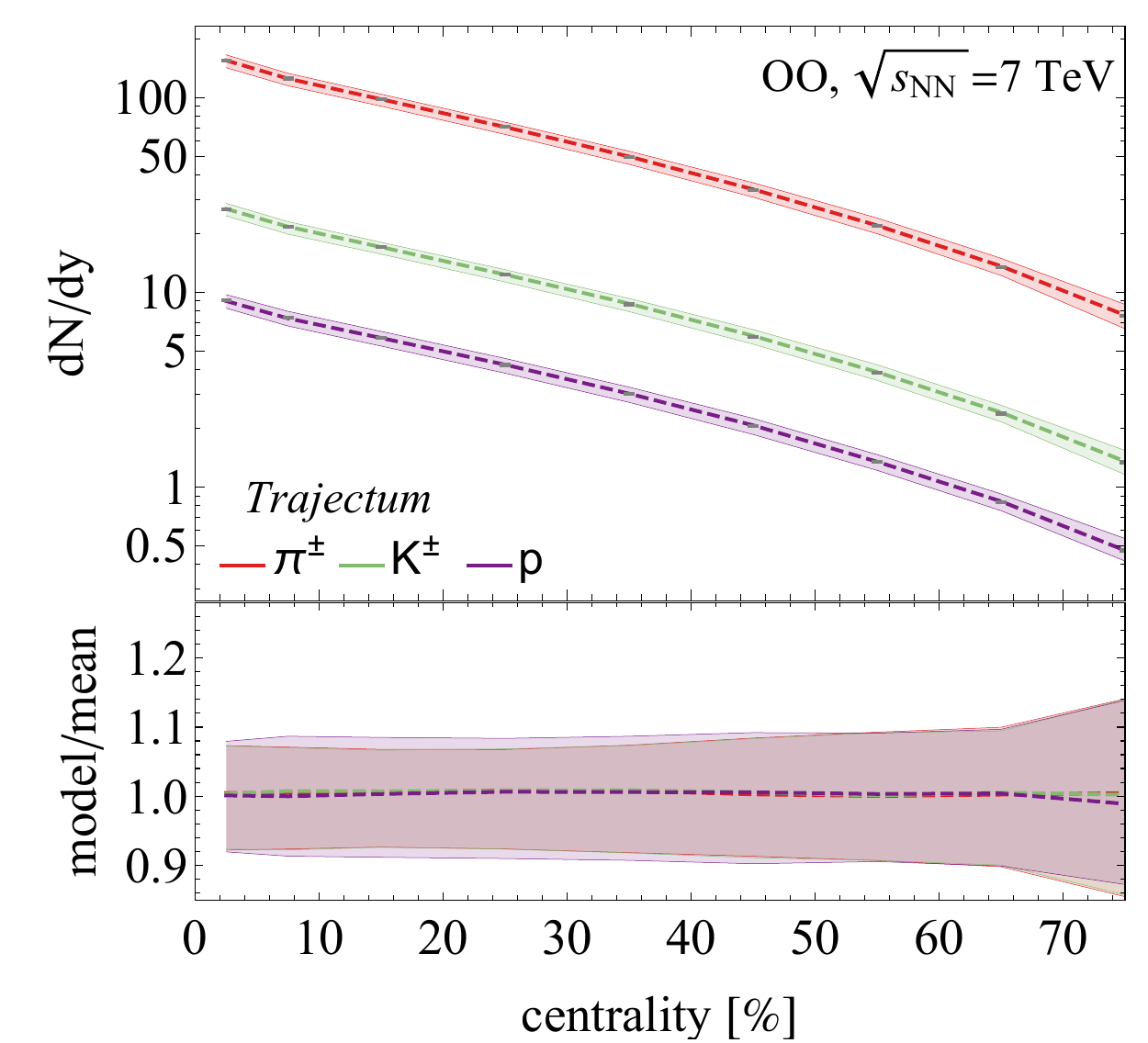}
\includegraphics[width=0.32\textwidth]{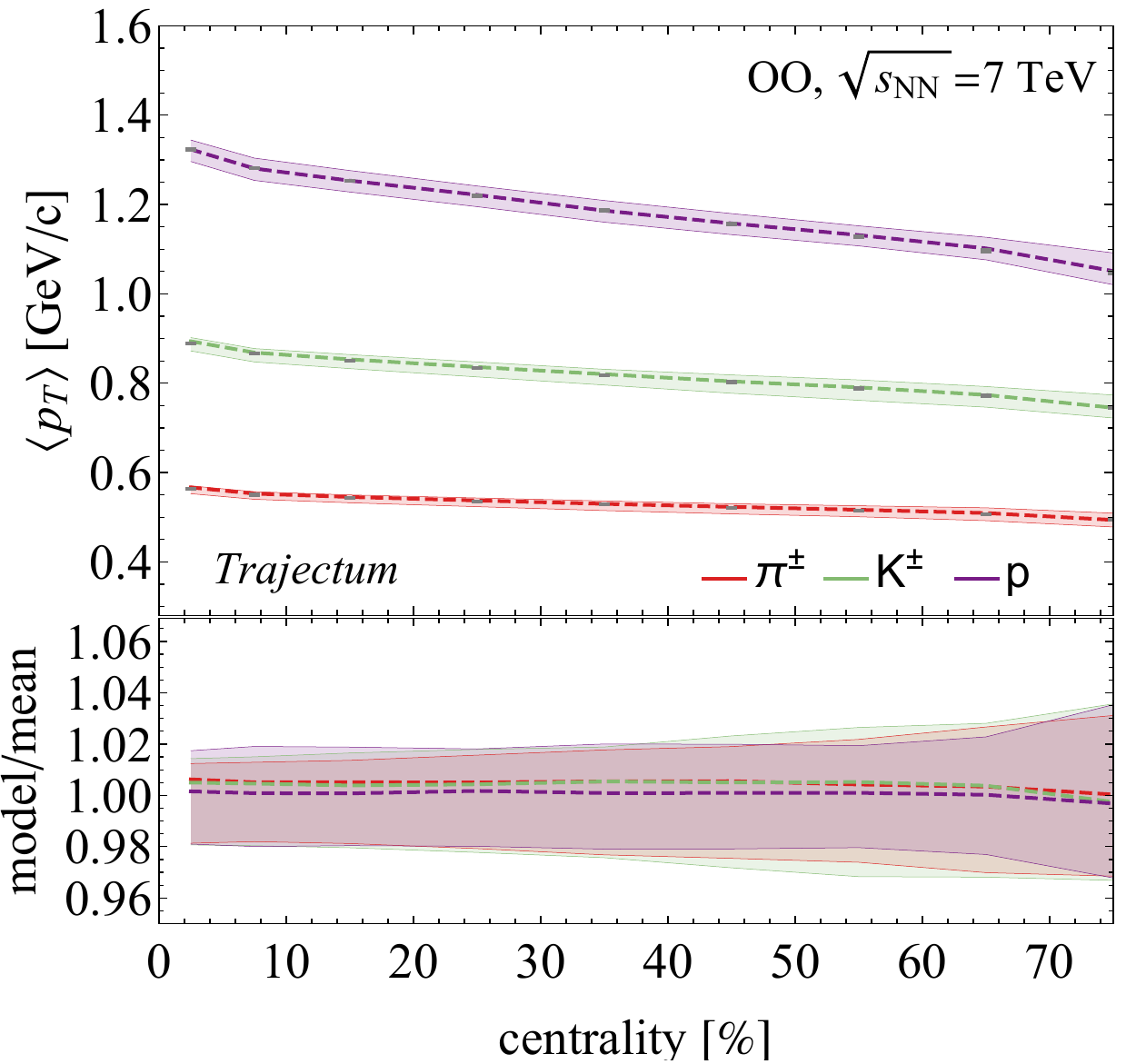}
\includegraphics[width=0.32\textwidth]{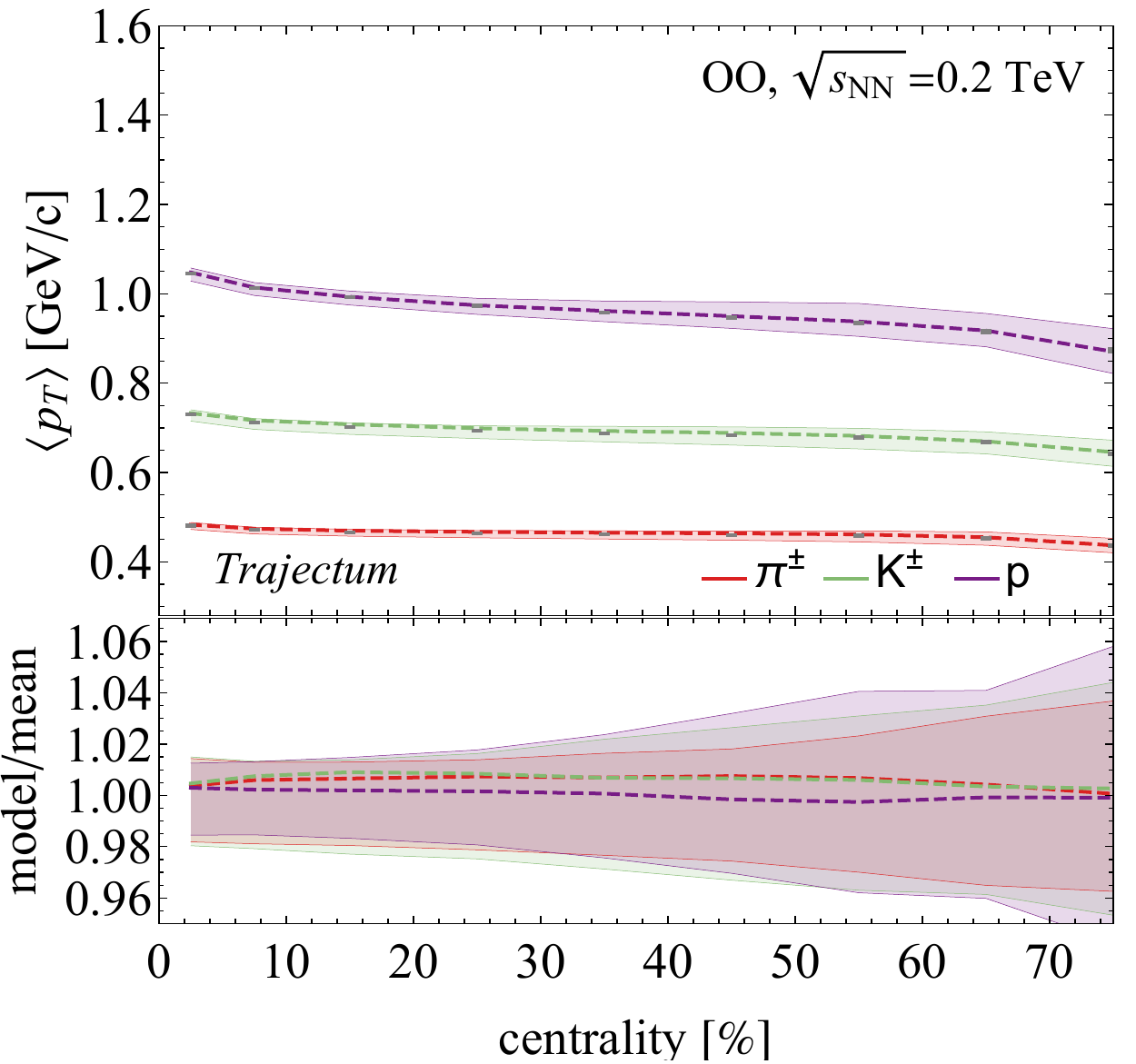}
\caption{\label{fig:multoo}We show identified particle yields $dN/dy$ (left) and mean $p_T$ (middle) for pions, kaons and protons at mid-rapidity as a function of centrality for OO collisions at $\sqrt{s_\text{NN}} = 7\,\text{TeV}$ and mean $p_T$ at $\sqrt{s_\text{NN}} = 0.2\,\text{TeV}$ (right)\@. The bands represent systematic uncertainties (1$\sigma$) in the \emph{Trajectum} parameters (see Fig.~\ref{fig:chainspbpb}) and the grey points show statistical uncertainties.}
\end{figure*}

The first such prediction is shown in Fig.~\ref{fig:mult}, where we show identified particle yields and mean transverse momenta for PbPb collisions at $\sqrt{s_\text{NN}} = 5.02\,\text{TeV}$ together with experimental results from ALICE \cite{ALICE:2019hno}\@. Note that as in \cite{Nijs:2020ors} at 5.02 TeV we only used charged particle multiplicity and integrated as well as $p_T$-differential $v_n\{k\}$ in the fit, so these particle identified multiplicities and $\mpt{}$ are true postdictions (the equivalent 2.76 TeV data was used in the fit however)\@.
The identified particle yields are consistent with the experimental data with the exception of the protons to which we will get back when we discuss the full spectra around Fig.~\ref{fig:spectra276}\@. The mean proton transverse momentum is above the experimental data, but still consistent within two standard deviations when taking into account our systematic uncertainty.

It is also interesting to examine observables which are currently not measured, as this allows us to make novel predictions. In Fig.~\ref{fig:multoo}, we show the same observables as in Fig.~\ref{fig:mult}, but for OO at $\sqrt{s_\text{NN}} = 7\,\text{TeV}$ and $0.2\,$TeV instead of PbPb at $\sqrt{s_\text{NN}} = 5.02\,\text{TeV}$\@. This is the (approximate) collision energy at which RHIC and the LHC collide the OO system. We are able to predict the identified particle yields to around 10\% precision, and the mean transverse momenta to around 2--3\%\@.

\begin{figure*}[ht]
\includegraphics[width=0.99\textwidth]{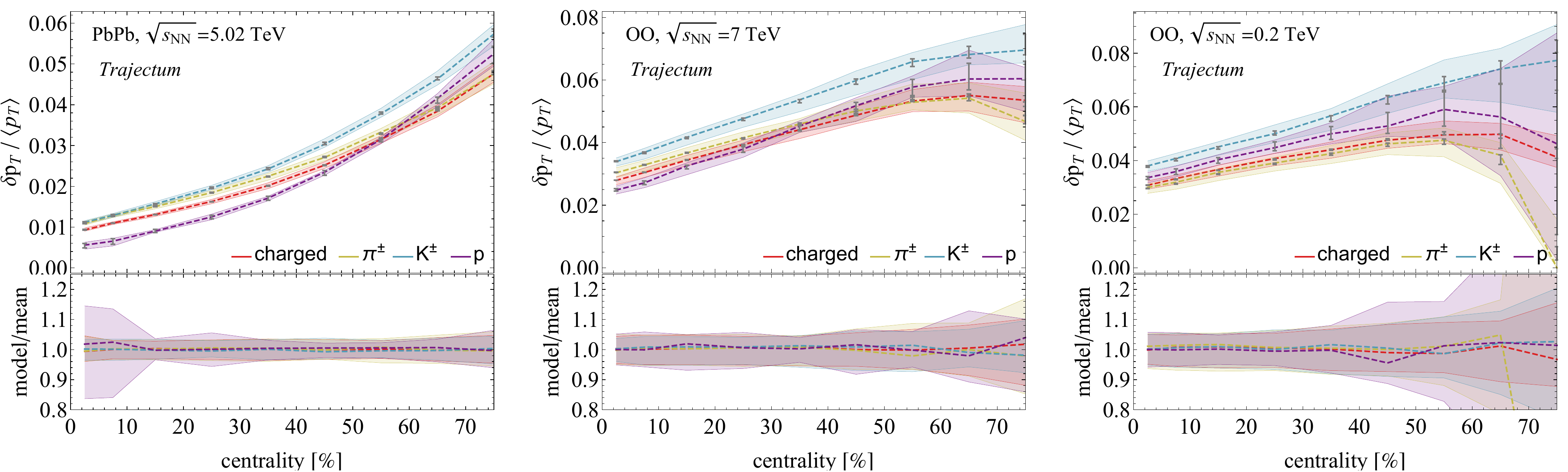}
\caption{\label{fig:ptfluct}We show transverse momentum fluctuations of charged particles, as well as pions, kaons and protons, all at mid-rapidity, as a function of centrality. This is done for PbPb collisions at $\sqrt{s_\text{NN}} = 5.02\,\text{TeV}$ (left), as well as OO collisions at $\sqrt{s_\text{NN}} = 7\,\text{TeV}$ (middle) and $\sqrt{s_\text{NN}} = 0.2\,\text{TeV}$ (right)\@. The bands represent systematic uncertainties (1$\sigma$) in the \emph{Trajectum} parameters (see Fig.~\ref{fig:chainspbpb}) and the grey points show statistical uncertainties.
}
\end{figure*}

We use transverse momentum fluctuations of charged particles in PbPb collisions at $\sqrt{s_\text{NN}} = 2.76\,\text{TeV}$ in the fit, but the experimental data for this observable is not available at $\sqrt{s_\text{NN}} = 5.02\,\text{TeV}$\@. In the left panel of Fig.~\ref{fig:ptfluct}, we show the \emph{Trajectum} prediction not only for charged particles, but also for pions, kaons and protons (see also \cite{Broniowski:2009fm, Giacalone:2020lbm, Schenke:2020uqq} for other predictions/postdictions)\@. For protons, the transverse momentum fluctuations are predicted with 5--15\% precision (statistics dominated, see the gray statistical uncertainties), depending on centrality, whereas the other three are predicted to less than 5\%\@. Curiously, the identified transverse momentum fluctuations do not seem to be ordered according to mass, but rather the kaon transverse momentum fluctuations are the largest. The middle and right panels of Fig.~\ref{fig:ptfluct} show the same observables, but this time for OO at $\sqrt{s_\text{NN}} = 7\,\text{TeV}$ (middle) and $\sqrt{s_\text{NN}} = 0.2\,\text{TeV}$ (right), which are the collision energies at which the LHC and RHIC, respectively, collide oxygen nuclei. One can see that the centrality dependence is weaker in OO from what it is in PbPb. Also, the uncertainties in OO are substantially larger than in PbPb, but still the precision for charged particles is less than 10\% for most centralities.

\begin{figure*}[ht]
\includegraphics[width=0.34\textwidth]{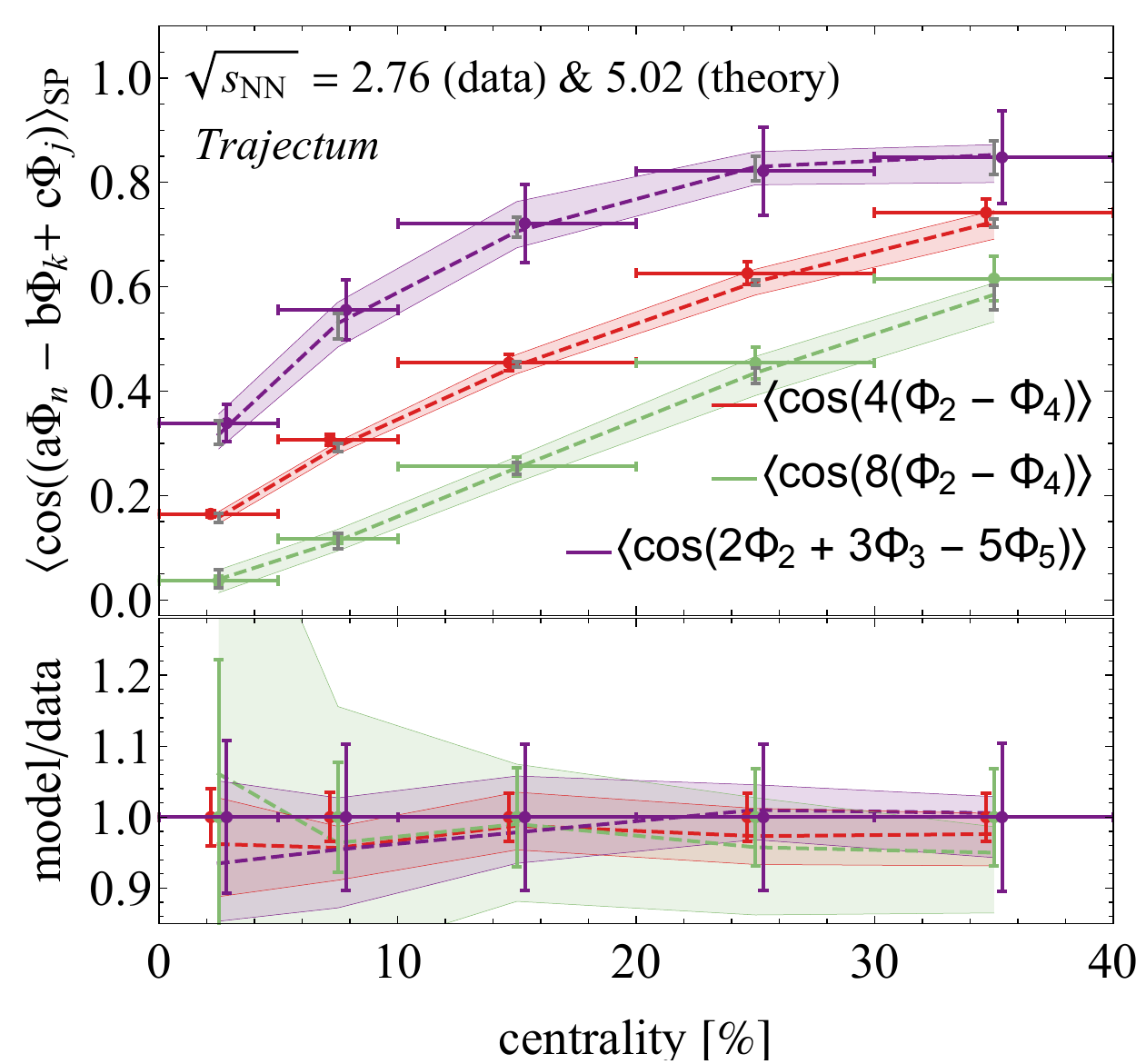}
\includegraphics[width=0.34\textwidth]{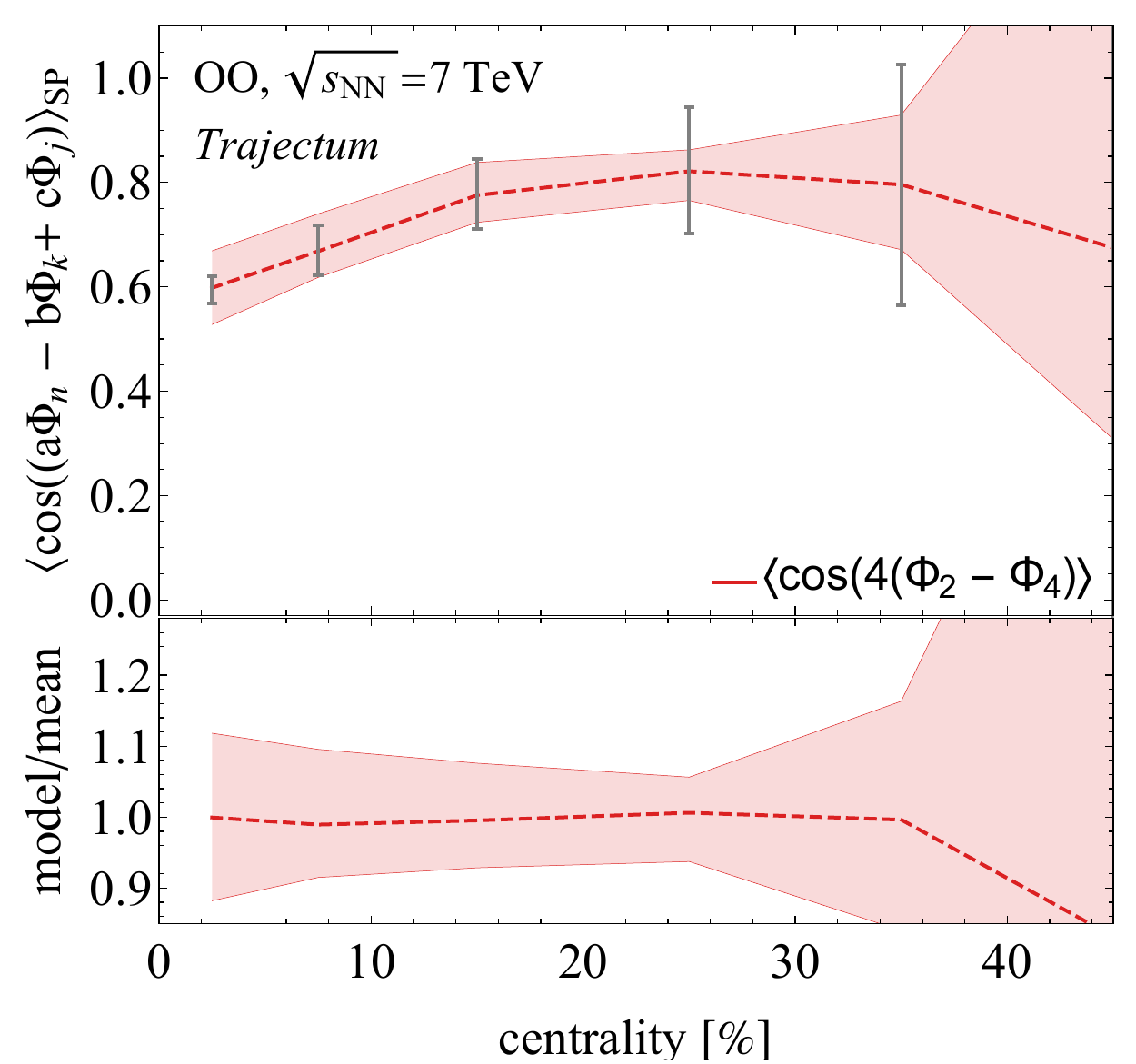}
\caption{\label{fig:psicor}We show event plane correlations $\langle\cos(4(\Phi_2 - \Phi_4))\rangle_\text{SP}$, $\langle\cos(8(\Phi_2 - \Phi_4))\rangle_\text{SP}$ and $\langle\cos(2\Phi_2 + 3\Phi_3 - 5\Phi_5)\rangle_\text{SP}$ of charged particles as a function of centrality for PbPb collisions at $\sqrt{s_\text{NN}} = 5.02\,\text{TeV}$ compared to colored ATLAS data points at $\sqrt{s_\text{NN}} = 2.76\,\text{TeV}$ (left) \cite{ATLAS:2014ndd}\@. We also show predictions for $\langle\cos(4(\Phi_2 - \Phi_4))\rangle_\text{SP}$ for OO at $\sqrt{s_\text{NN}} = 7\,\text{TeV}$ (right)\@. The bands represent systematic uncertainties (1$\sigma$) in the \emph{Trajectum} parameters (see Fig.~\ref{fig:chainspbpb}) and the grey points show statistical uncertainties.}
\end{figure*}

An interesting family of observables which we did not use in the fit are the event plane angle correlations. In Fig.~\ref{fig:psicor}, we show $\langle\cos(4(\Phi_2 - \Phi_4))\rangle_\text{SP}$, $\langle\cos(8(\Phi_2 - \Phi_4))\rangle_\text{SP}$ and $\langle\cos(2\Phi_2 + 3\Phi_3 - 5\Phi_5)\rangle_\text{SP}$ for PbPb at $\sqrt{s_\text{NN}} = 5.02\,\text{TeV}$ in the left panel. We compare the theory prediction to data at $\sqrt{s_\text{NN}} = 2.76\,\text{TeV}$ because this measurement is not available at $\sqrt{s_\text{NN}} = 5.02\,\text{TeV}$\@.
What makes these observables interesting is that they are not obviously related to the observables which we used in the fit, but we still describe them with an accuracy that can be even smaller than the experimental uncertainty. This means that, at least within our model, the event plane angle correlations are related to the observables that we do use in the fit in a non-trivial way, so that they can be completely described in terms of these simpler observables. In the right panel of Fig.~\ref{fig:psicor}, we make a prediction for $\langle\cos(4(\Phi_2 - \Phi_4))\rangle_\text{SP}$ for OO at $\sqrt{s_\text{NN}} = 7\,\text{TeV}$\@. For centrality bins up to about 30\% centrality we are able to make a prediction with a precision of about 10\%, with precision deteriorating for more peripheral collisions due to our limited statistics.

\begin{figure*}[ht]
$\vcenter{\hbox{\includegraphics[width=0.33\textwidth]{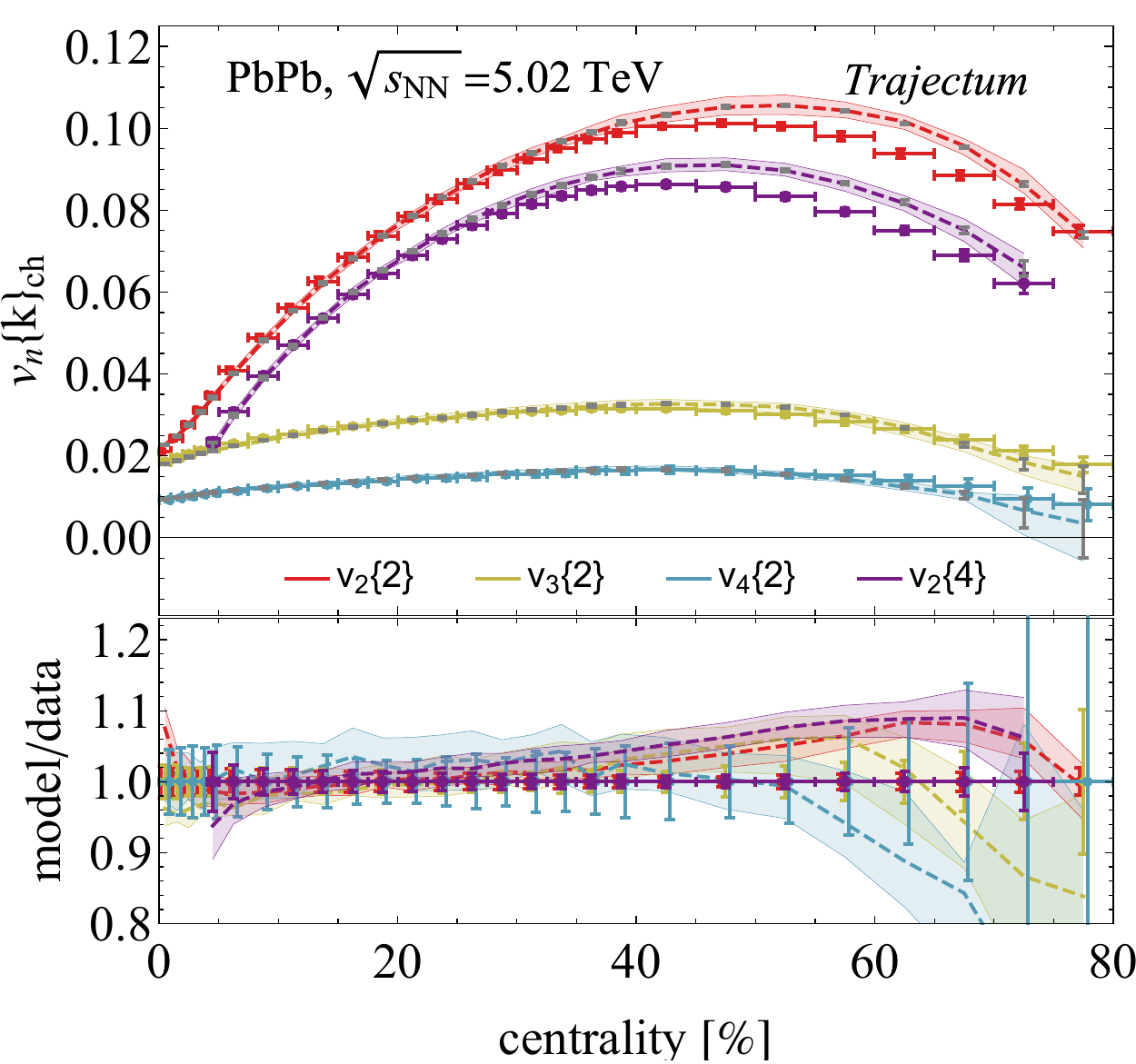}}}$
$\vcenter{\hbox{\includegraphics[width=0.33\textwidth]{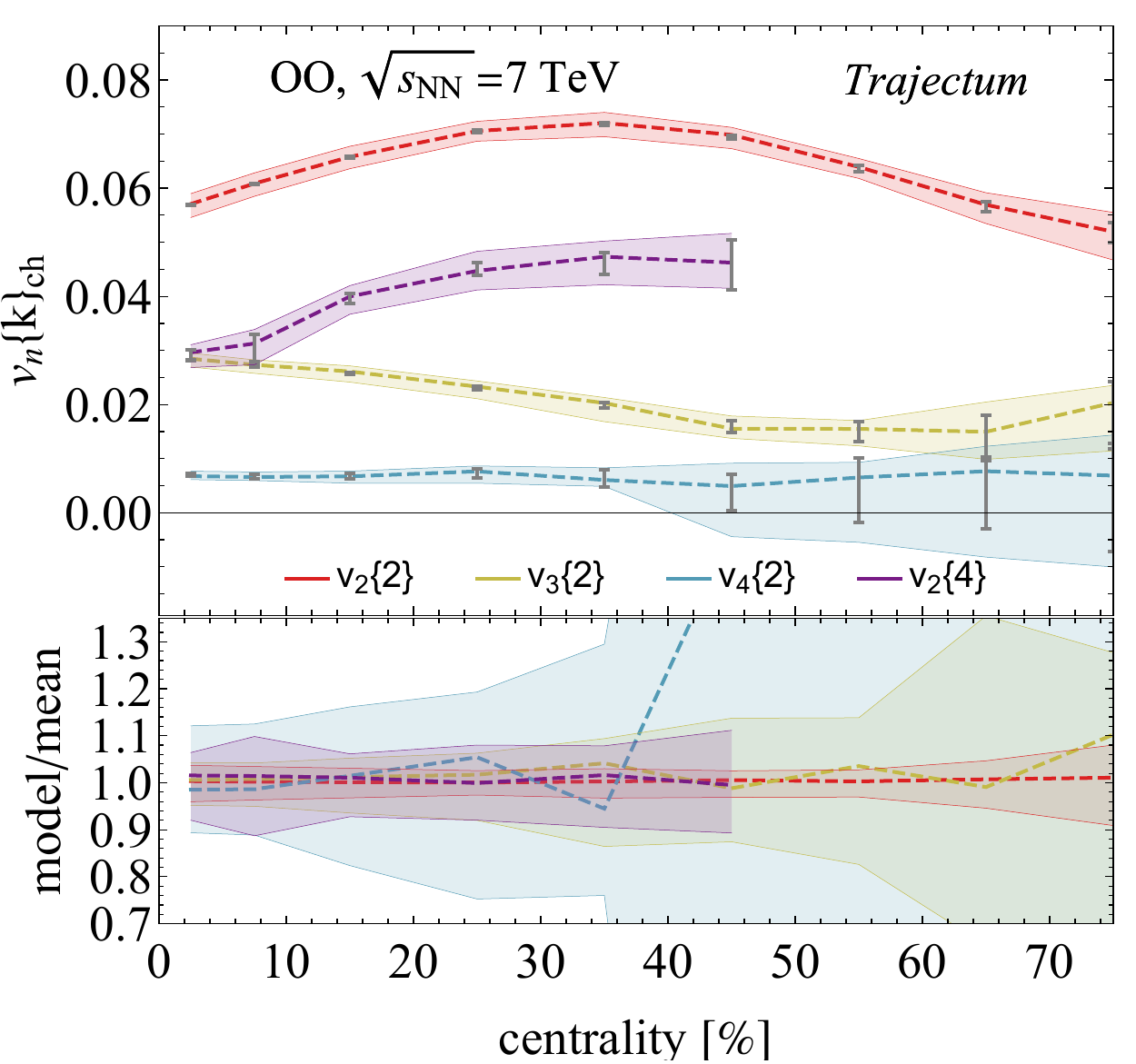}}}$
$\vcenter{\hbox{\includegraphics[width=0.32\textwidth]{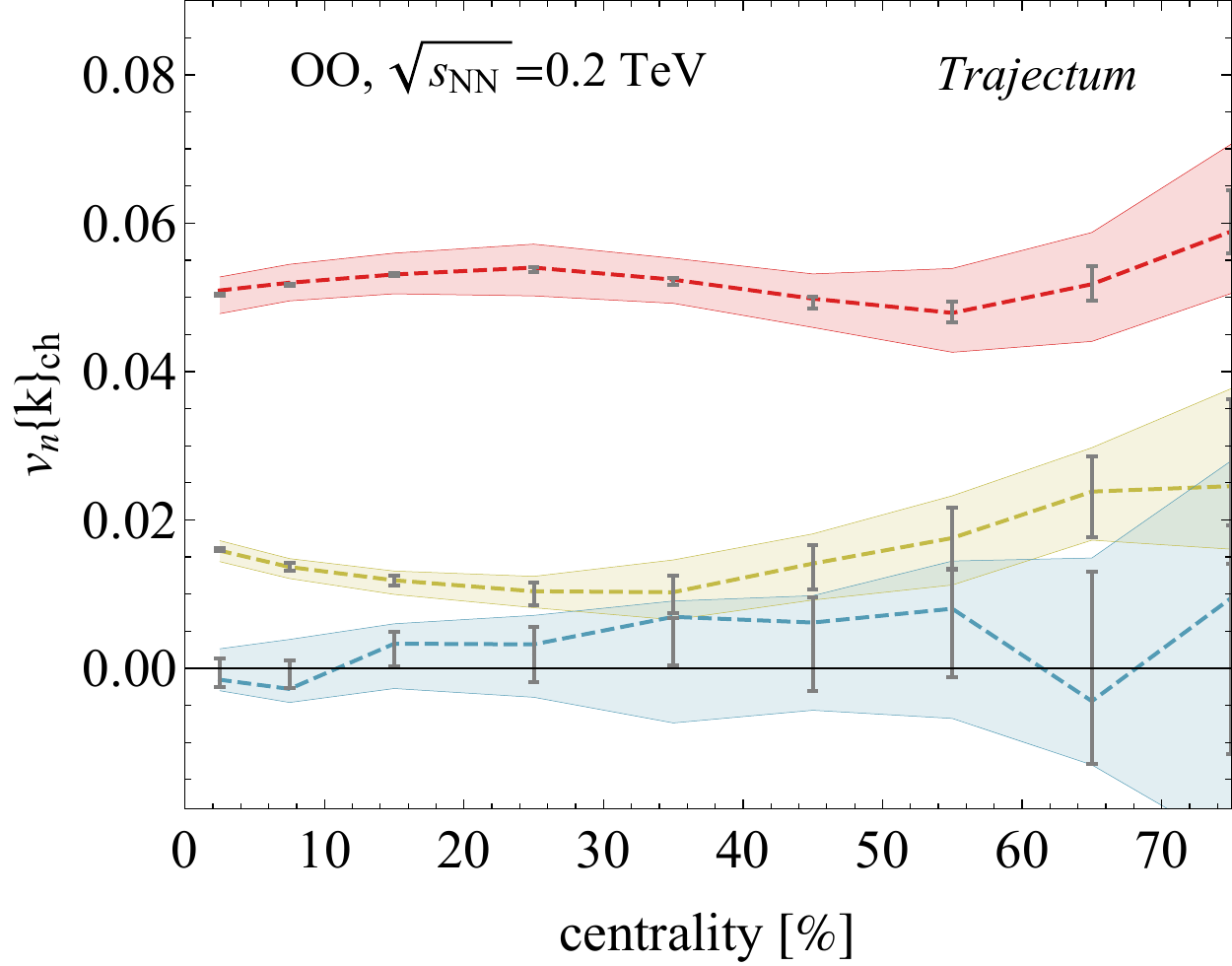}}}$
\caption{\label{fig:vnchargedint}We show integrated anisotropic flow coefficients $v_n\{k\}$ of charged particles at mid-rapidity as a function of centrality for PbPb collisions at $\sqrt{s_\text{NN}}= 5.02\,$TeV compared to colored ALICE data points \cite{ALICE:2018rtz} (left), predictions for OO collisions at 7 TeV (middle) and 0.2 TeV (right). The bands represent systematic uncertainties (1$\sigma$) in the \emph{Trajectum} parameters (see Fig.~\ref{fig:chainspbpb}) and the grey points show statistical uncertainties.}
\end{figure*}

\begin{figure*}[ht]
\includegraphics[width=0.99\textwidth]{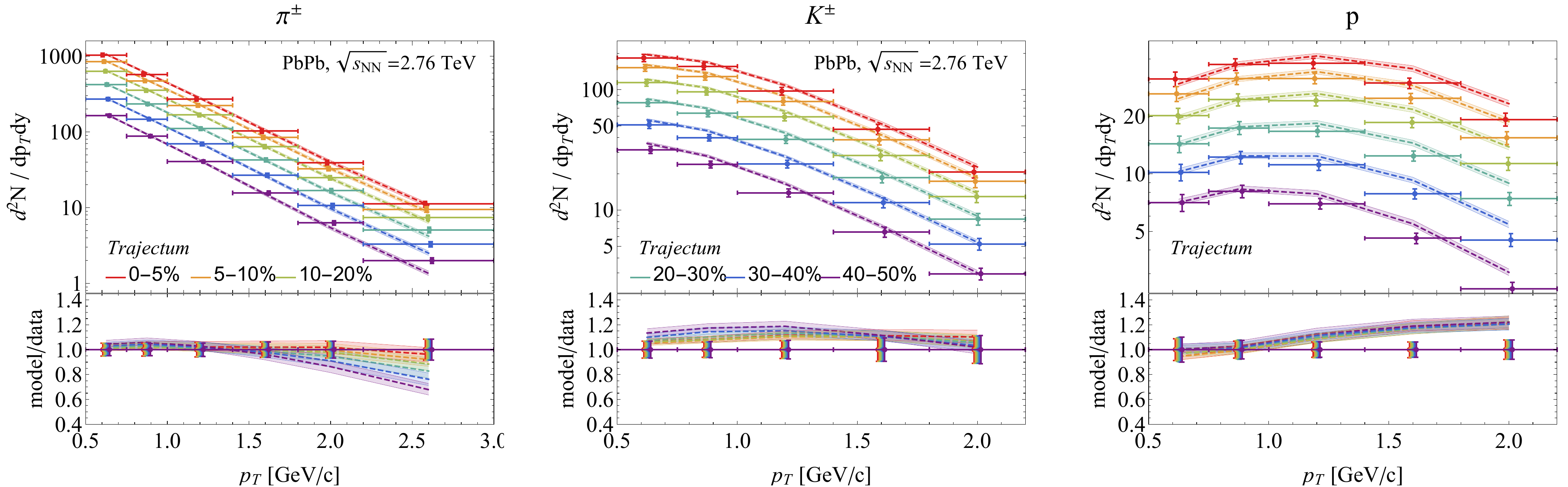}
\caption{\label{fig:spectra276}We show the identified transverse momentum spectra for pions (left), kaons (middle) and protons (right) at mid-rapidity for different centralities for PbPb collisions at $\sqrt{s_\text{NN}} = 2.76\,\text{TeV}$ compared to colored ALICE data points \cite{Abelev:2013vea}\@. The bands represent systematic uncertainties (1$\sigma$) in the \emph{Trajectum} parameters (see Fig.~\ref{fig:chainspbpb})\@. We note as opposed to our 5.02 TeV results these spectra are generated using the emulator and hence also contain an uncertainty due to the emulator.
}
\end{figure*}

In the fit, we have used integrated anisotropic flow coefficients $v_n\{k\}$ at both $\sqrt{s_\text{NN}} = 2.76\,\text{TeV}$ and $\sqrt{s_\text{NN}} = 5.02\,\text{TeV}$, with centrality bins bounded by $(0, 5, 10, 20, 30, 40, 50, 60, 70)$\%, statistics allowing.
In the left panel of Fig.~\ref{fig:vnchargedint}, we show the \emph{Trajectum} results for $\sqrt{s_\text{NN}} = 5.02\,\text{TeV}$ using finer centrality bins.
One can see that we obtain good precision, around 5--10\% for most centrality bins.
For most centralities, we also have excellent agreement with ALICE data, with the largest deviations being around 2$\sigma$ in $v_2\{2\}$ and $v_2\{4\}$ for centralities from 50\% to 70\%\@.
In the middle and right panels, we again show predictions for OO collisions at $\sqrt{s_\text{NN}} = 7\,\text{TeV}$ (middle) and $\sqrt{s_\text{NN}} = 0.2\,\text{TeV}$ (right)\@.
We obtain predictions with a precision of less than 10\% for most flow coefficients, with $v_3\{2\}$ and $v_4\{2\}$ being less precisely predicted for peripheral collisions.
This can be understood, however, as $v_3\{2\}$ becomes very small for peripheral collisions, while $v_4\{2\}$ is very small for every centrality class.
Also note that the systematic uncertainties in all three panels include values where $v_n\{k\} < 0$, whereas from its definition $v_n\{k\}$ is either positive real or the $k$-th root of a negative number\@.
To make sense of the complex values, we redefine $v_n\{k\}$ as $\tilde v_n\{k\} = \sgn(v_n\{k\}^k)|v_n\{k\}|$\@.
This essentially rotates the complex values to the negative real axis while keeping the real ones unchanged, making $\tilde v_n\{k\}$ real in all cases.
In Fig.~\ref{fig:vnchargedint}, wherever $v_n\{k\}$ becomes complex, we therefore show $\tilde v_n\{k\}$ instead. We note that $v_2\{4\}$ in PbPb collisions also becomes complex for centralities smaller than 4\% \cite{ATLAS:2019peb} (see also \cite{Nijs:2020roc}), but our current simulation does not have sufficient statistics to fully probe this regime.

\begin{figure*}[ht]
\includegraphics[width=0.99\textwidth]{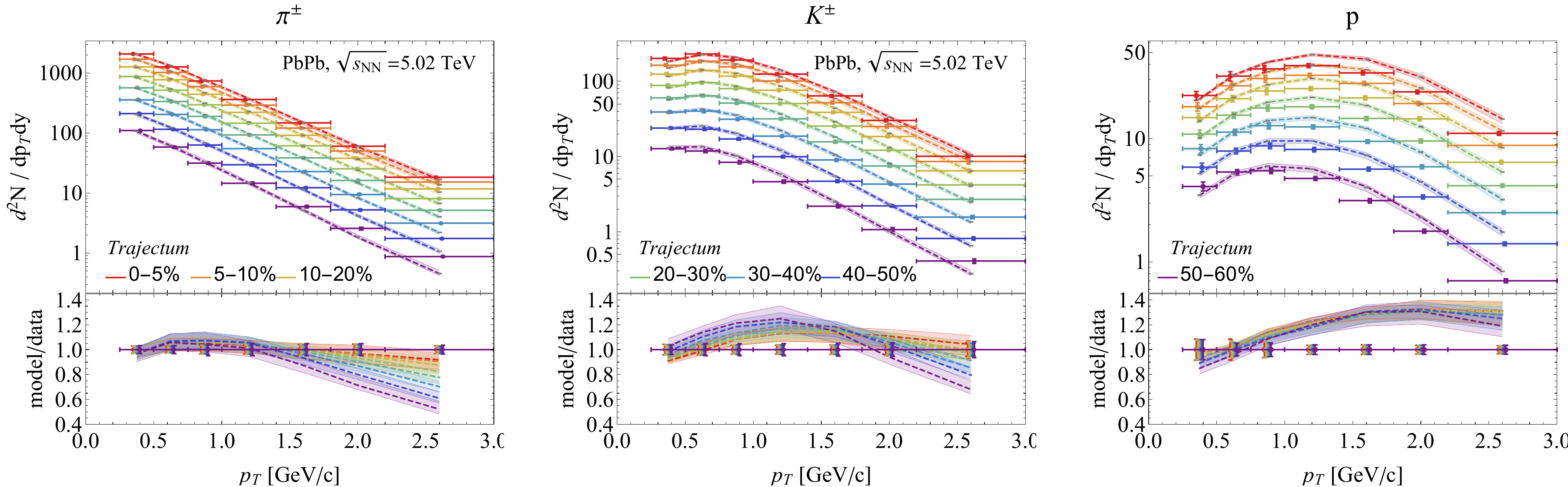}
\caption{\label{fig:spectra}We show the identified transverse momentum spectra for pions (left), kaons (middle) and protons (right) at mid-rapidity for different centralities for PbPb collisions at $\sqrt{s_\text{NN}} = 5.02\,\text{TeV}$ compared to colored ALICE data points \cite{ALICE:2019hno}\@. The bands represent systematic uncertainties (1$\sigma$) in the \emph{Trajectum} parameters (see Fig.~\ref{fig:chainspbpb}) and the grey points show statistical uncertainties.
}
\end{figure*}

\begin{figure*}[ht]
\includegraphics[width=0.99\textwidth]{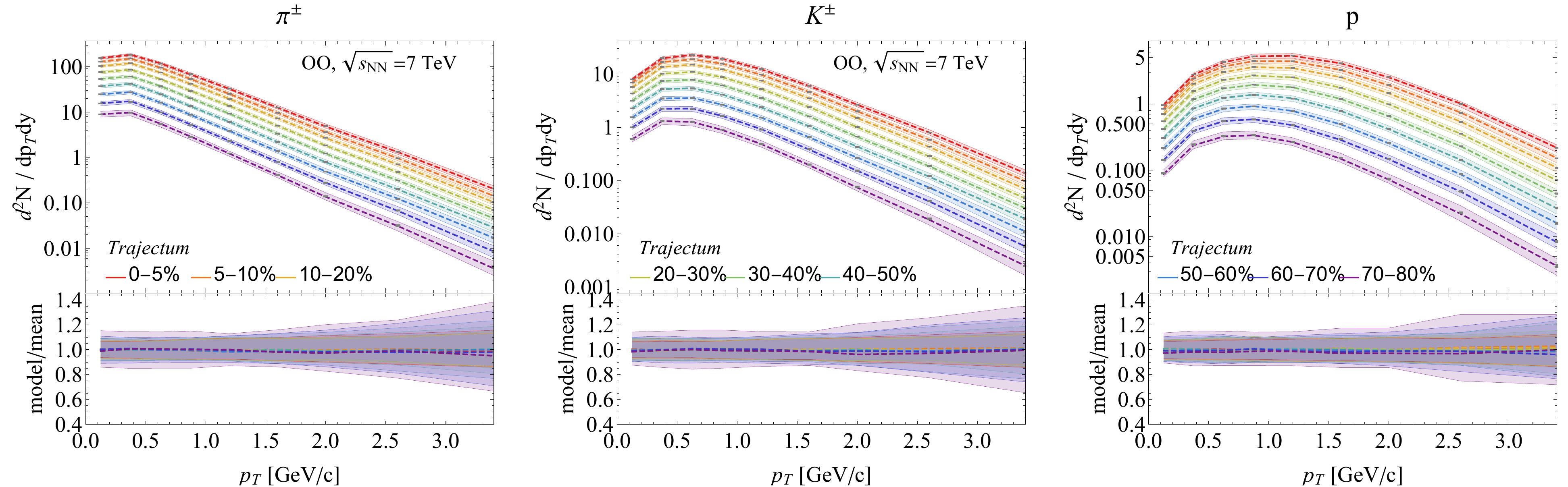}
\includegraphics[width=0.99\textwidth]{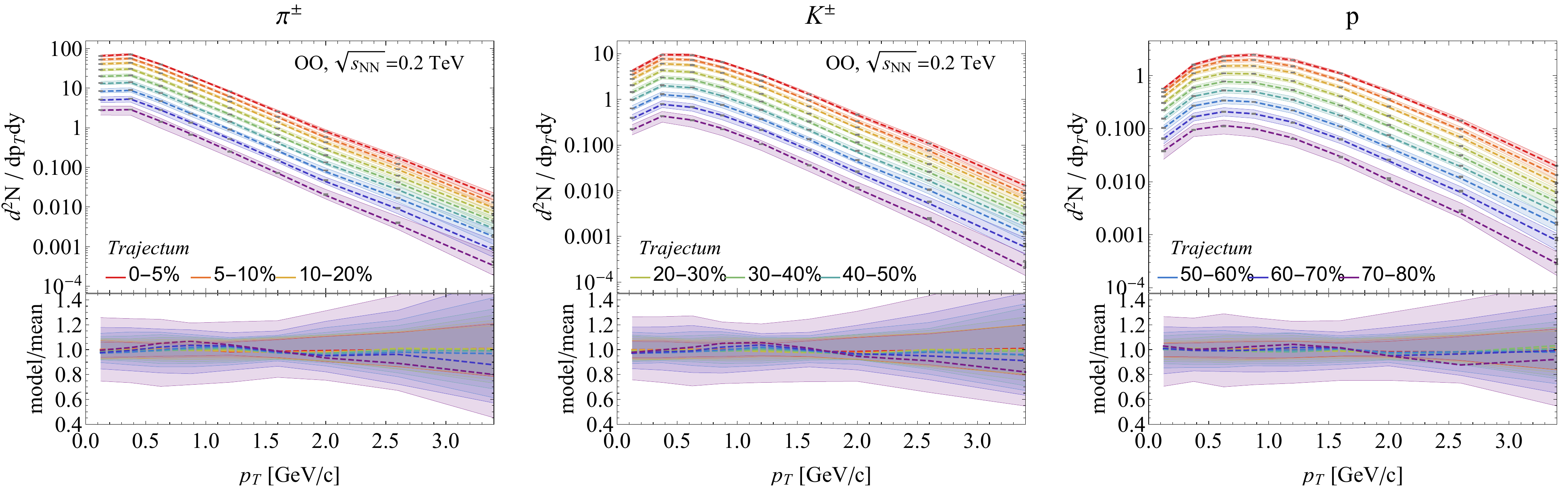}
\caption{\label{fig:spectraoo}We show the identified transverse momentum spectra for pions (left), kaons (middle) and protons (right) at mid-rapidity for different centralities for OO collisions at $\sqrt{s_\text{NN}} = 7\,\text{TeV}$ (top) and $\sqrt{s_\text{NN}} = 0.2\,\text{TeV}$ (bottom)\@. The bands represent systematic uncertainties (1$\sigma$) in the \emph{Trajectum} parameters (see Fig.~\ref{fig:chainspbpb}) and the grey points show statistical uncertainties.
}
\end{figure*}

\subsection{$p_T$-differential predictions}

In addition to the various $p_T$-integrated observables discussed above, we also compute several $p_T$-differential observables. In Fig.~\ref{fig:spectra276}, we show identified transverse momentum spectra for pions (left), kaons (middle) and protons (right), for PbPb at $\sqrt{s_\text{NN}} = 2.76\,\text{TeV}$\@. Note that, in contrast to the other figures described in this section, the contents of Fig.~\ref{fig:spectra276} were not made using a high statistics run, but instead come from evaluating the emulator which was used to perform the Bayesian analysis. Whereas deviations from the model with the experimental data is typically small (around 10\%), it is clear that especially for more off-central collisions we have too few high $p_T$ pions. This could indicate that perturbative QCD processes (which are only power-law suppressed at high $p_T$ as opposed to the exponential Boltzmann suppression) become more important towards more peripheral collisions. Also mostly at peripheral collisions we have too many kaons as well as high $p_T$ protons. 

Fig.~\ref{fig:spectra} shows the same quantities as Fig.~\ref{fig:spectra276}, but now for PbPb at $\sqrt{s_\text{NN}} = 5.02\,\text{TeV}$\@. In this figure, the high statistics run was used to provide the \emph{Trajectum} prediction. Again the precision is around 10\%, but here deviations from the experimental data are much larger, especially as the experimental uncertainty has decreased significantly with respect to 2.76 TeV data. Given that we did not fit to the spectra at $\sqrt{s_\text{NN}} = 5.02\,\text{TeV}$, this is not too alarming, but it does indicate that including these measurements in a future fit is likely to improve constraints. We also show predictions for identified transverse momentum spectra for OO in Fig.~\ref{fig:spectraoo}, where we show pions (left), kaons (middle) and protons (right) at $\sqrt{s_\text{NN}} = 7\,\text{TeV}$ (top) and $\sqrt{s_\text{NN}} = 0.2\,\text{TeV}$ (bottom)\@. The precision of the predictions depends on centrality and collision energy, but is around 10--30\% for most centrality and $p_T$-bins.

\begin{figure*}[ht!]
\includegraphics[width=0.99\textwidth]{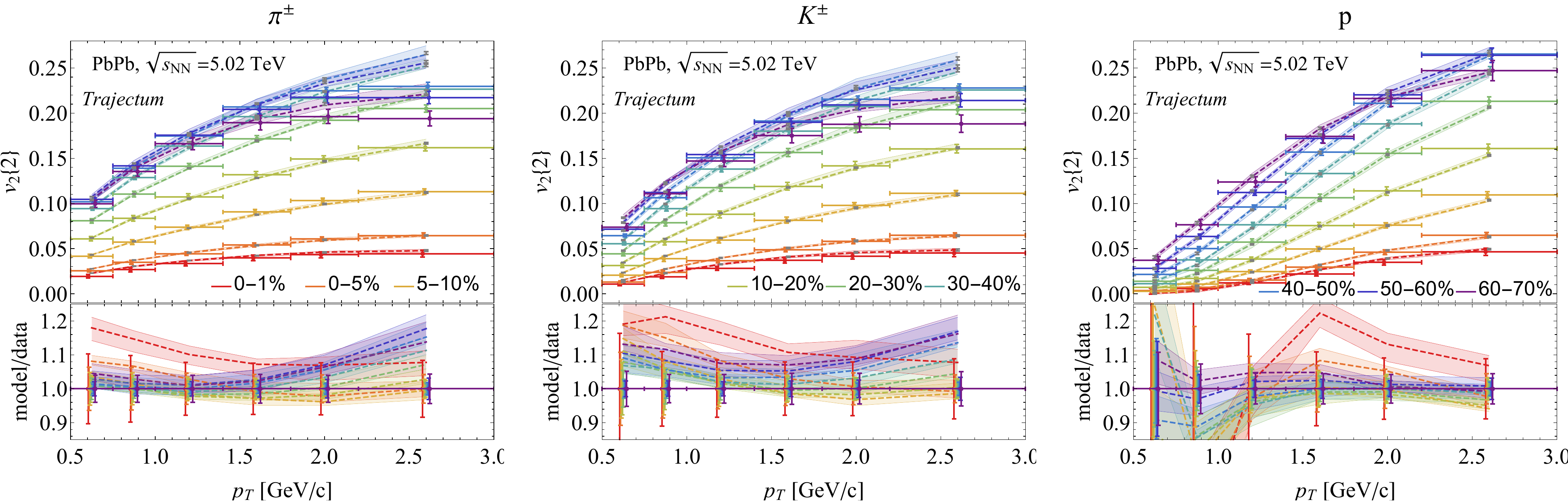}
\caption{\label{fig:vnkptdiff}We show the identified $p_T$-differential anisotropic flow coefficient $v_2\{2\}(p_T)$ for pions (left), kaons (middle) and protons (right) at mid-rapidity for different centralities for PbPb collisions at $\sqrt{s_\text{NN}} = 5.02\,\text{TeV}$ compared to colored ALICE data points \cite{ALICE:2018yph}\@. The bands represent systematic uncertainties (1$\sigma$) in the \emph{Trajectum} parameters (see Fig.~\ref{fig:chainspbpb}) and the grey points show statistical uncertainties.
}
\end{figure*}

Lastly, Fig.~\ref{fig:vnkptdiff} shows the identified $p_T$-differential anisotropic flow coefficient $v_2\{2\}(p_T)$ for pions (left), kaons (middle) and protons (right), for PbPb at $\sqrt{s_\text{NN}} = 5.02\,\text{TeV}$\@. Depending on centrality and transverse momentum, the precision ranges from around 5 to 10\%\@. The largest relative deviations from experimental data can be seen in the 0--1\% centrality bin (this particular centrality bin was not used in the fit), but in absolute terms these still agree within two standard deviations. The only serious deviations are the overestimations of the elliptic flow for high $p_T$ pions and kaons for centralities bigger than 40\%. This could again indicate the importance of pQCD contributions, as those contributions are expected to decrease the flow coefficients $v_n$\@. The agreement of the proton $v_2\{2\}$ is however impressively precise over the entire range of centralities (up to 70\%) and transverse momentum probed.

\vspace{30 mm}

\section{Predictions for ultracentral collisions}%
\label{sec:v2v3puzzle}

\begin{figure}[ht]
\includegraphics[width=0.36\textwidth]{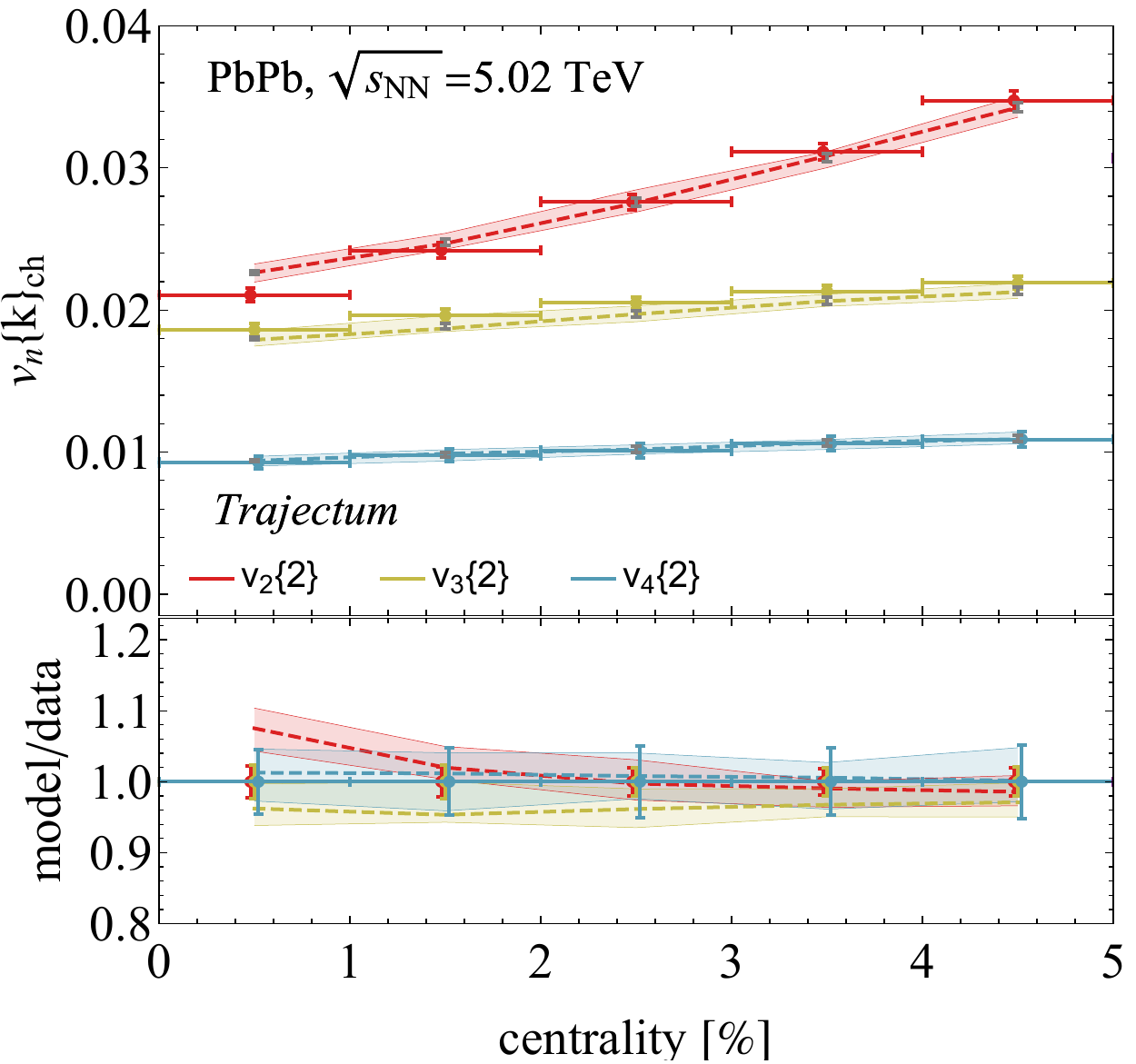}
\caption{\label{fig:vnchargedintcentral}We show a zoom of Fig.~\ref{fig:vnchargedint} (left) together with data from ALICE \cite{ALICE:2018rtz}\@.
}
\end{figure}

\begin{figure*}[ht]
\includegraphics[width=0.32\textwidth]{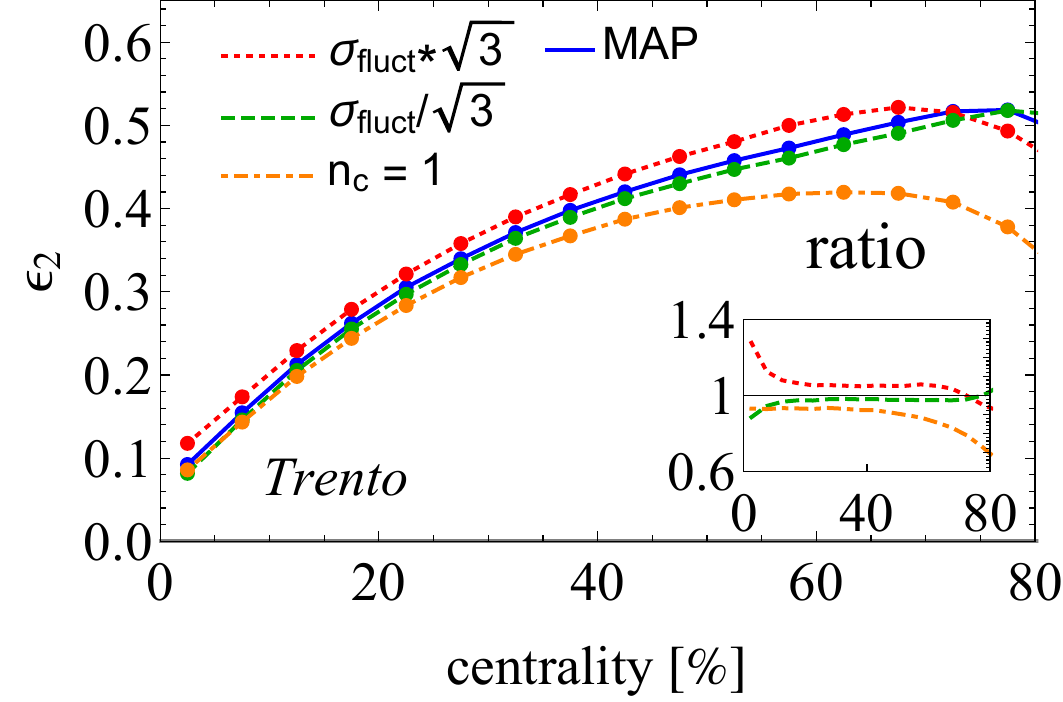}
\includegraphics[width=0.32\textwidth]{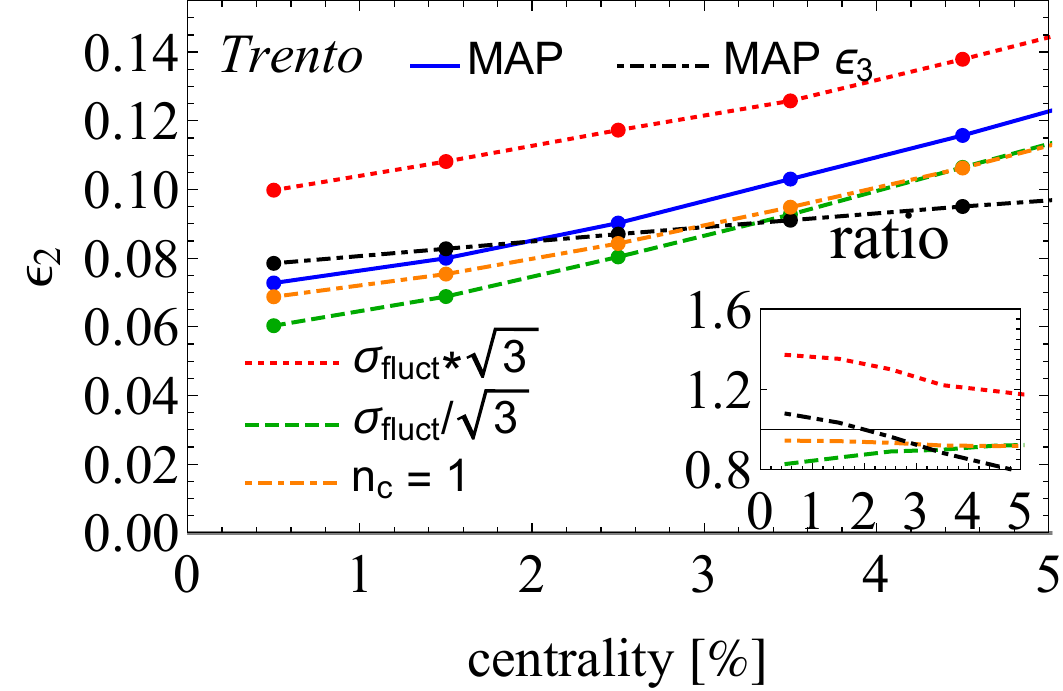}
\includegraphics[width=0.32\textwidth]{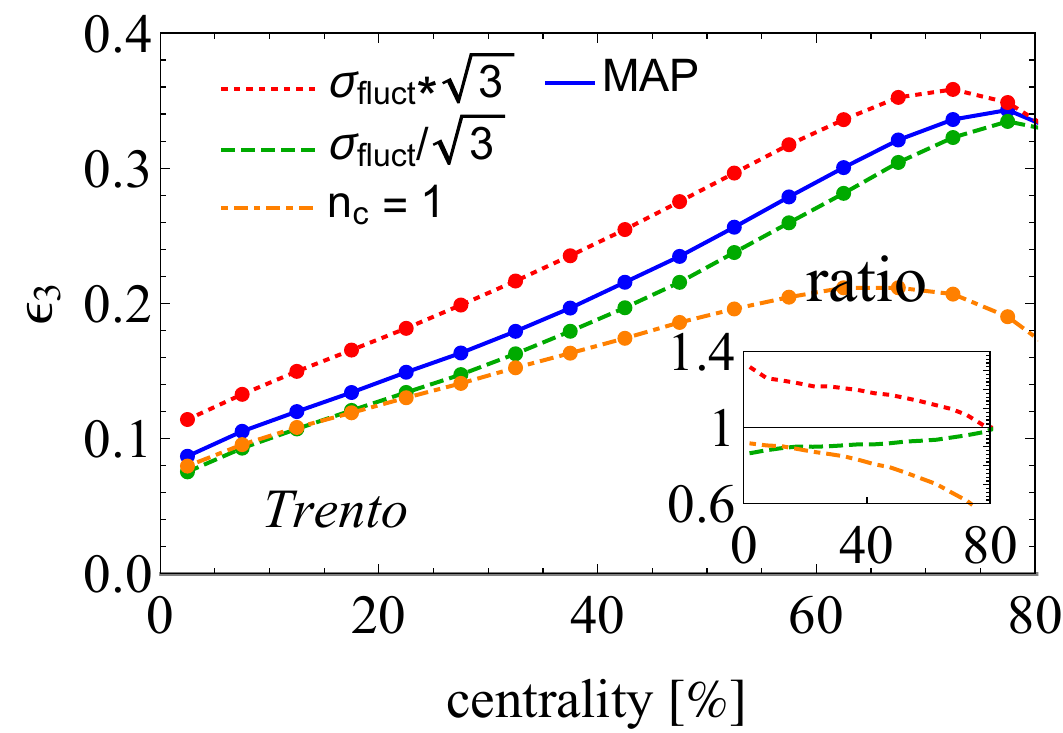}
\caption{\label{fig:trento}We show a simple \trento{} simulation of the initial eccentricities $\epsilon_2$ (left, middle) and $\epsilon_3$ (right), where in the middle panel we show a zoom of the panel on the left together with $\epsilon_3$ for comparison (black dot-dashed). In addition to the maximum a posteriori (MAP) values, we show results where $\sigma_\text{fluct}$ was rescaled by a factor $\sqrt{3}$, as well as results where the number of constituents $n_c$ was set to 1\@. As insets we also show ratios of these variations with respect to the MAP settings. One can see that $\sigma_\text{fluct}$ has a large influence on both $\epsilon_2$ and $\epsilon_3$, which is part of the resolution of the ultracentral $v_2$ to $v_3$ puzzle \cite{CMS:2012xxa,Gelis:2019vzt,Carzon:2020xwp,Snyder:2020rdy}\@.}
\end{figure*}

\begin{figure*}[ht]
\includegraphics[width=0.3\textwidth]{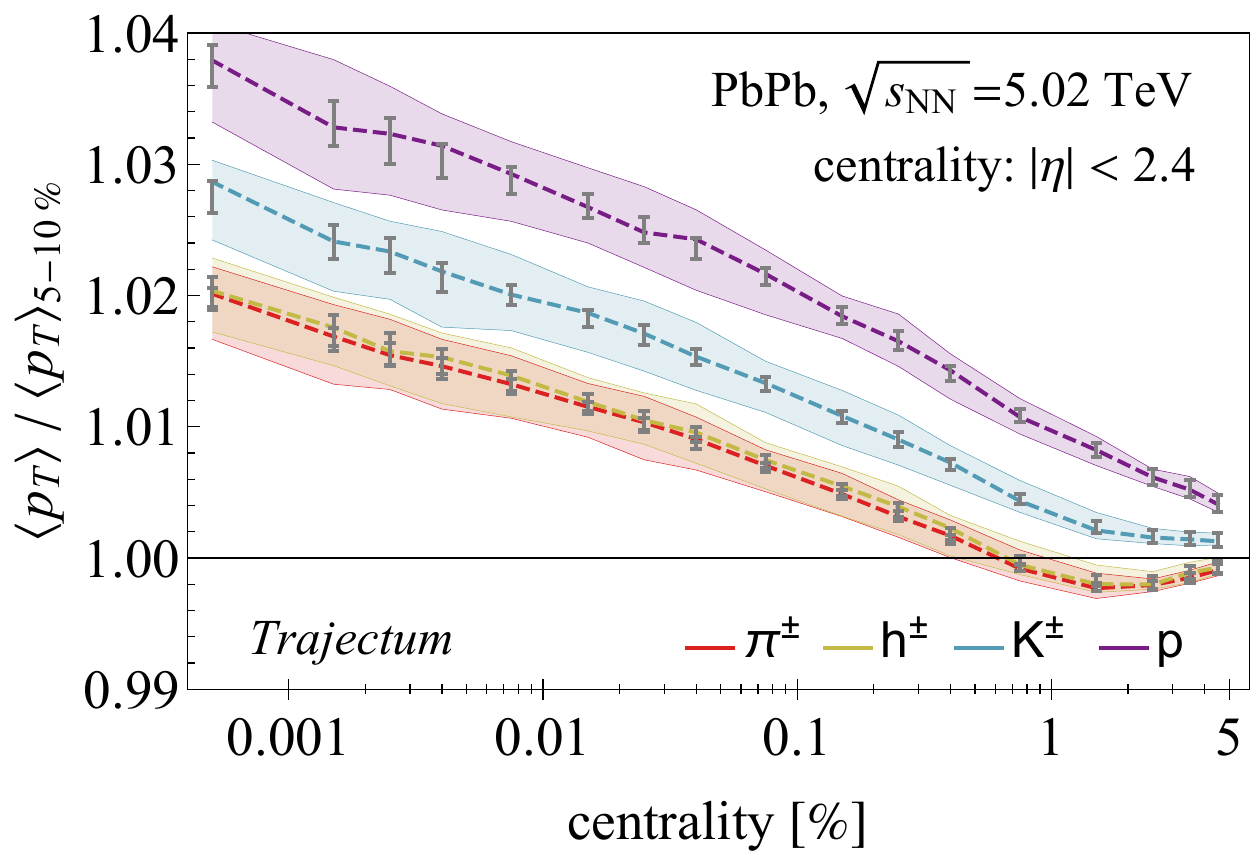}
\includegraphics[width=0.3\textwidth]{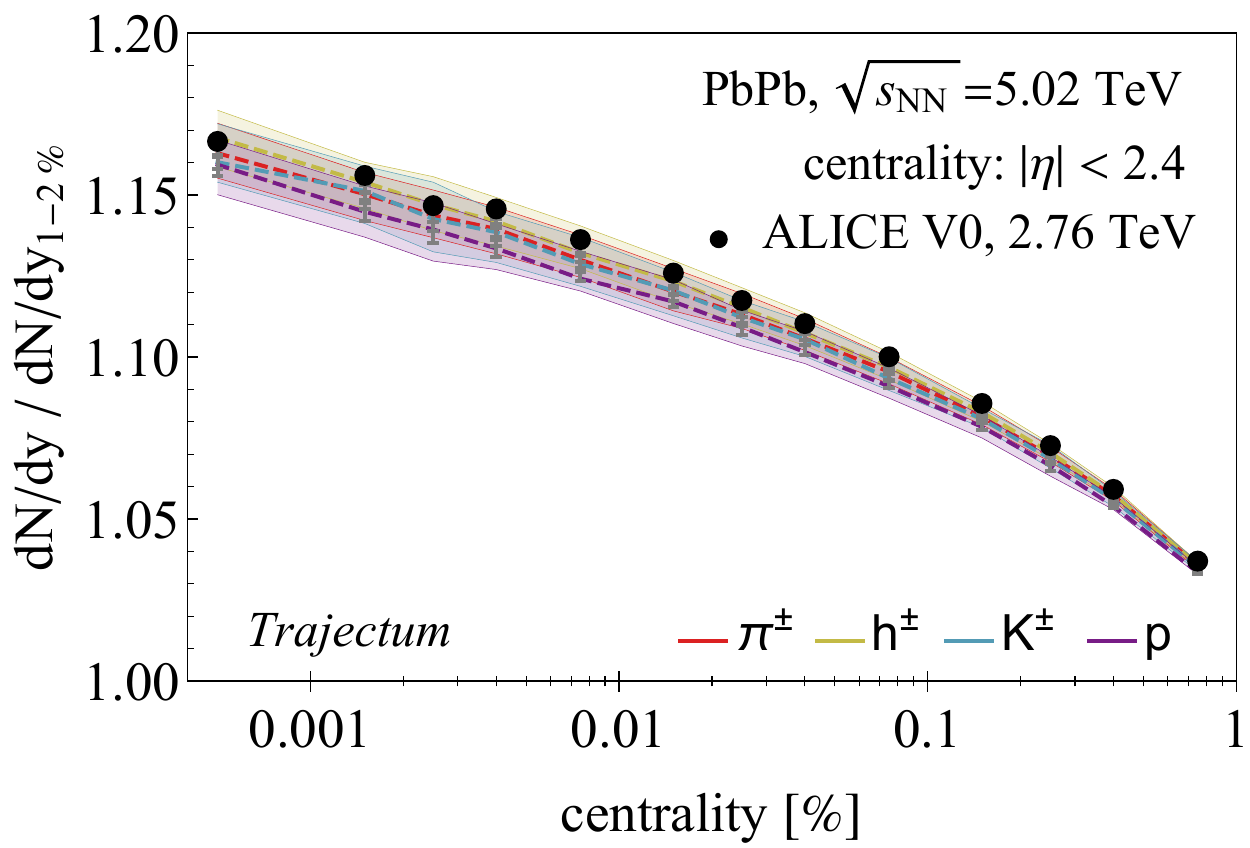}
\includegraphics[width=0.3\textwidth]{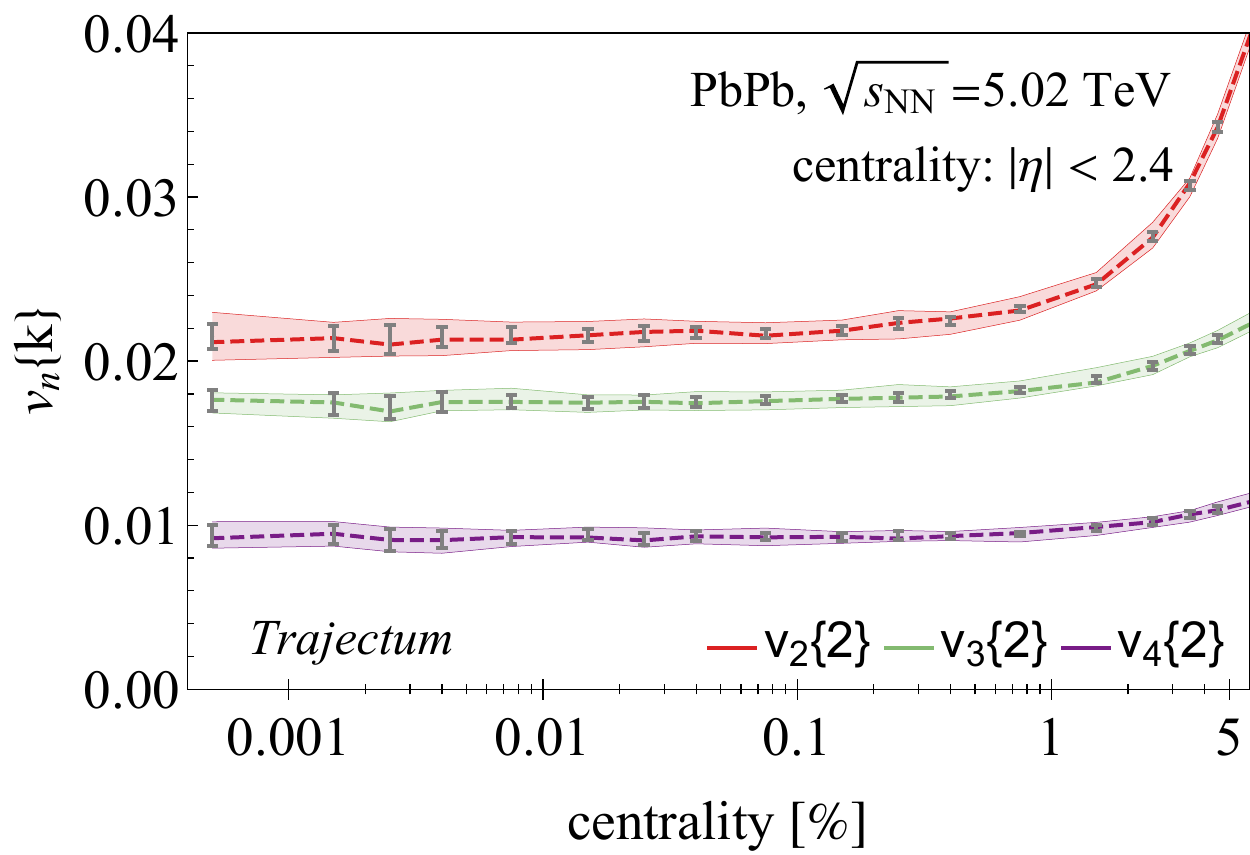}
\includegraphics[width=0.3\textwidth]{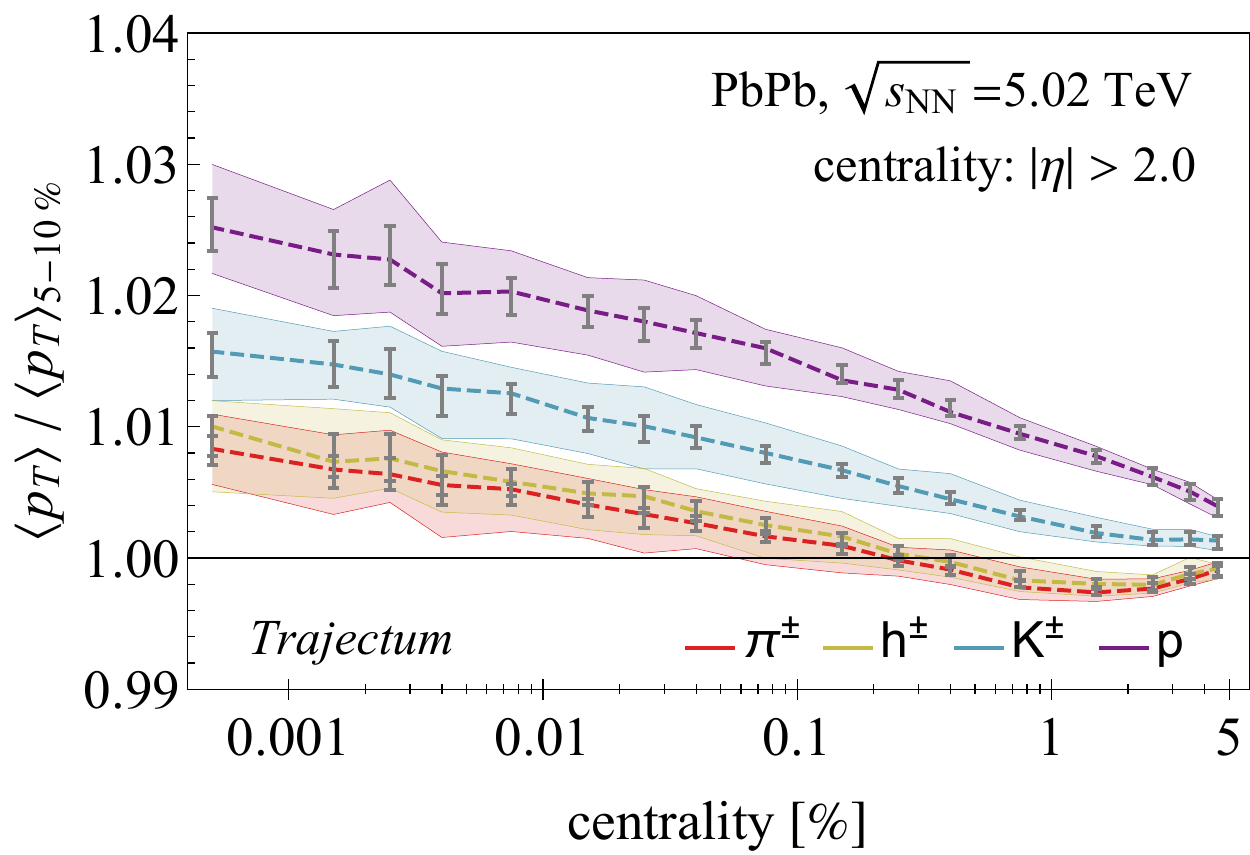}
\includegraphics[width=0.3\textwidth]{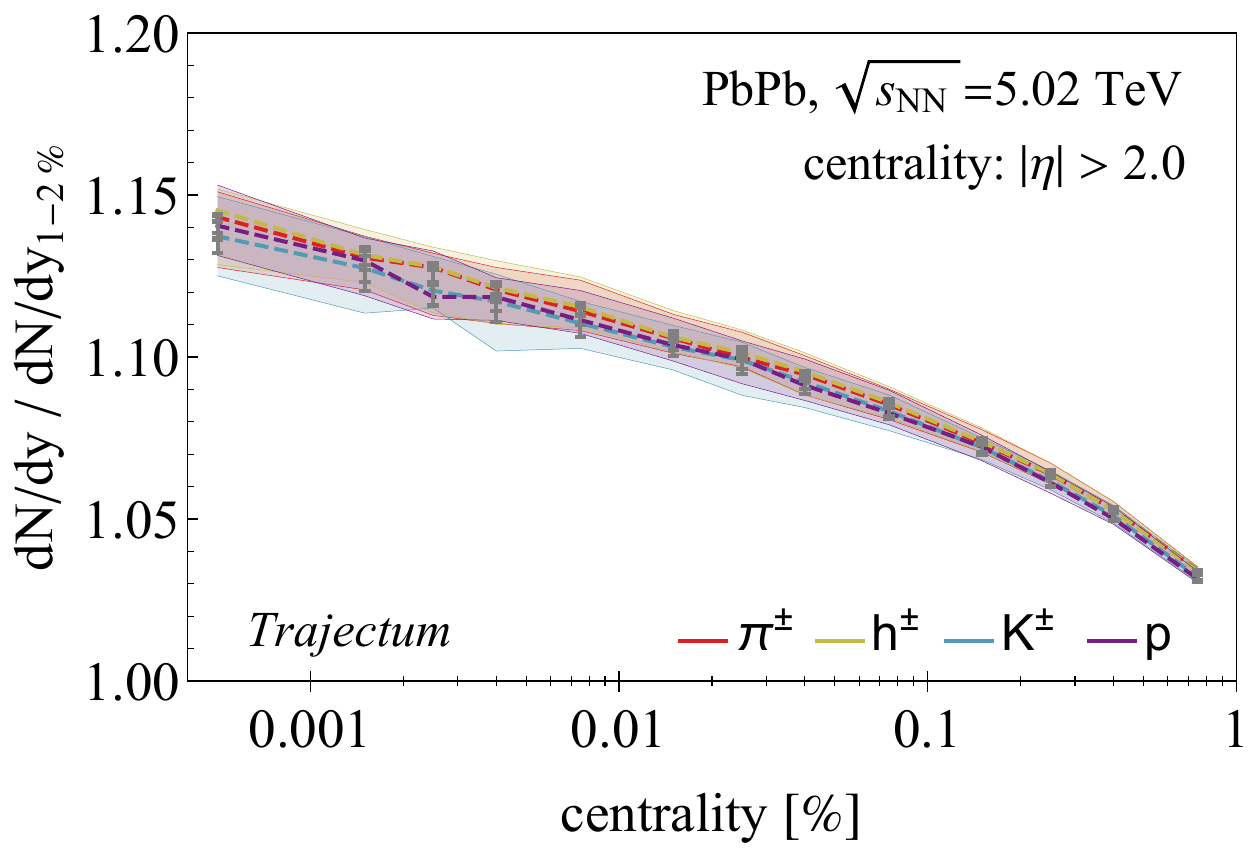}
\includegraphics[width=0.3\textwidth]{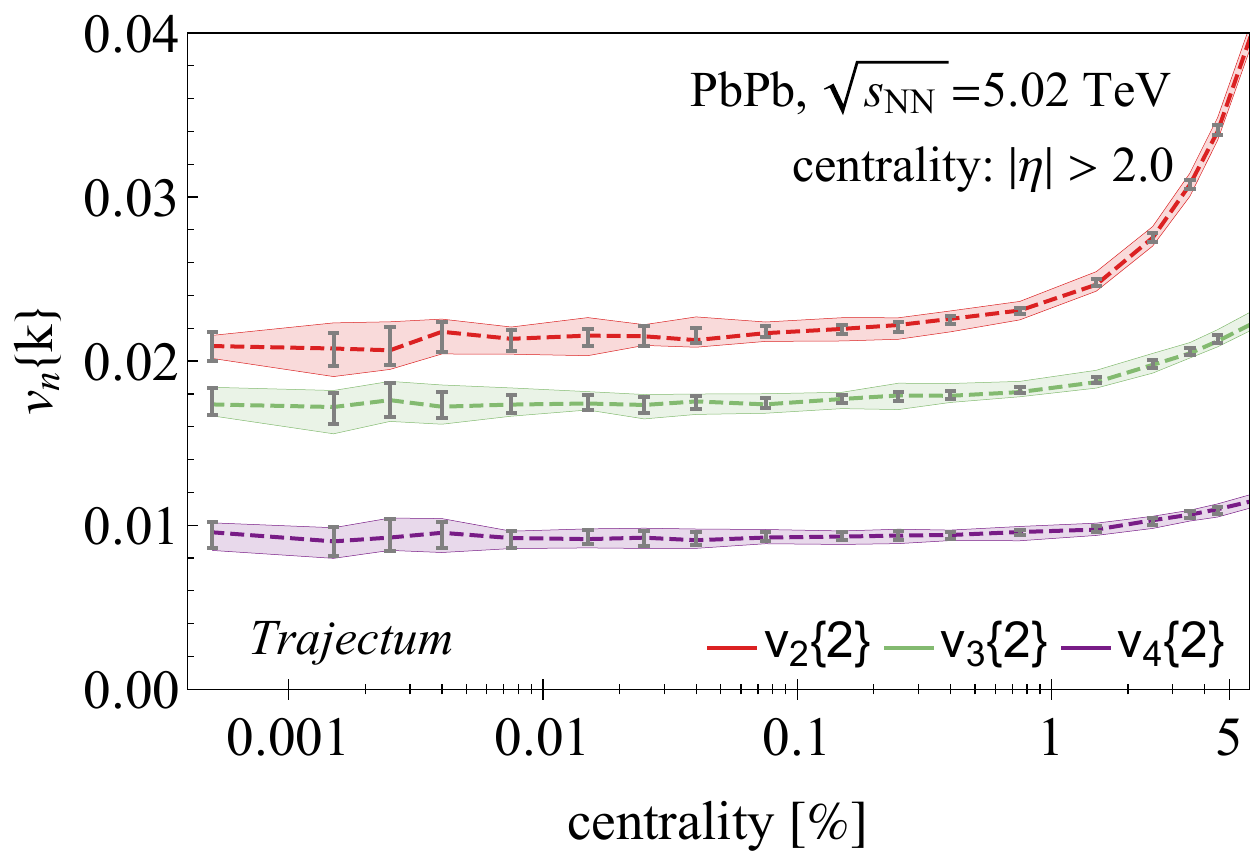}

\caption{\label{fig:ultracentral}(top) We show the mean transverse momentum for charged pions, hadrons, kaons and protons normalized to its value in the 0--5\% centrality class towards extremely ultracentral collisions (left, see also \cite{Gardim:2019brr}) and similarly for the multiplicity at mid-rapidity (middle, we use $dN/d\eta$ for unidentified hadrons) and the $v_n\{2\}$ coefficients (right, not normalized)\@. While the 3\% difference for protons for the mean transverse momentum naively amounts to an effect that is within the systematic uncertainty the normalization with the 0--5\% class should allow much of the systematics to cancel, such that with the high luminosity delivered to ALICE this measurement is likely feasible. The $v_n\{2\}$ do not decrease further towards extremely ultracentral collisions, indicating that the QGP never becomes perfectly spherical.
(bottom) Instead of our standard centrality measure (charged multiplicity with $p_T > 400\,\text{MeV}$ and $|\eta|<2.4$ we show the same figures for a $|\eta|>2.0$ cut, which is similar to the forward V0 centrality definition typically used by ALICE\@. By construction the effect on the multiplicity is now smaller (since we select in a different rapidity range), but an even greater reduction is found for the $\langle p_T \rangle$\@. Even bigger effects from the centrality measure can be expected for models that do not assume boost invariance.
}
\end{figure*}

Ultracentral collisions in the 0--1\% centrality class are rare events whereby the nuclei collide head-on and furthermore fluctuate specifically to have a high multiplicity. These collisions hence probe a special part of heavy ion collision physics, and have been historically difficult to describe accurately. A relatively famous example is the $v_2$-$v_3$-ultracentral puzzle, which states that it is difficult to simultaneously describe both $v_2$ and $v_3$ in the 0--1\% centrality class \cite{CMS:2012xxa,Gelis:2019vzt,Carzon:2020xwp,Snyder:2020rdy}, but also for mean transverse momentum there are theoretical predictions \cite{Gardim:2019brr}\@. The interesting observation in the latter work is that for extremely ultracentral collisions (0--0.01\% centrality) the size of the plasma is limited by the radius of the colliding nucleus and the increase in multiplicity is solely sourced by an upward fluctuation in the initial temperature. The increase in temperature increases the final radial flow and hence leads to an increase in $\langle p_T \rangle$\@.

In Fig.~\ref{fig:vnchargedintcentral} we show \tr{} postdictions for $v_n\{k\}$ as compared to ALICE measurements \cite{ALICE:2018rtz} (this Figure is a zoom of Fig.~\ref{fig:vnchargedint} (left))\@. This yields an excellent fit to the experimental data \cite{ALICE:2018rtz}, even though these particular centrality classes were not used in the Bayesian model. The $v_2$-$v_3$-puzzle is usually stated as a puzzle on the ratio, since it is often straightforward to reduce or increase the magnitude of both by modifying the shear viscosity. This, however, would modify the $v_n\{k\}$ for higher centrality classes and we use our parameters within the posterior range that are constrained by such data. In that case it can be challenging to fit the absolute magnitude of both $v_2$ and $v_3$ (indeed \cite{Carzon:2020xwp} is off by many standard deviations for $v_2\{2\}$, while the $v_2\{2\}/v_3\{2\}$ ratio is consistent with data within a single standard deviation)\@. We remark that both \cite{Schenke:2020mbo} and \cite{Bernhard:2018hnz,Bernhard:2019bmu} have an excellent description of the 0--5\% centrality class of both $v_2\{2\}$ and $v_3\{2\}$ so we regard it as likely that they will also describe the ultracentral centrality classes. The good fit of the Magma model \cite{Gelis:2019vzt} however likely does not survive a full hydrodynamic simulation \cite{Snyder:2020rdy}\@.

It is an interesting question why \tr{} would obtain a better ultracentral description than e.g.~v-USPhydro in \cite{Carzon:2020xwp}, especially as the latter work uses the same initial state model (\trento{}) and furthermore attempted some important changes to resolve the puzzle (albeit unsuccessfully)\@. For this question we performed a simple \trento{} simulation \cite{Moreland:2014oya, Moreland:2018gsh} using our MAP settings from \cite{Nijs:2020ors} to obtain the spatial anisotropies $\epsilon_2$ and $\epsilon_3$, here defined by the initial energy density through \cite{Moreland:2014oya}
\begin{equation}
\epsilon_n = \left|\frac{\int r\,dr\,d\phi\,r^n e^{i n \phi} e}{\int r\,dr\,d\phi\,r^n e}\right|.
\end{equation}
We furthermore show three variations obtained by setting the number of constituents to $n_c=1$ as well as rescaling $\sigma_\text{fluct}$ by a factor $\sqrt{3}$\@. We note that $\epsilon_2$ and $\epsilon_3$ correlate very well with $v_2\{2\}$ and $v_3\{2\}$, respectively \cite{Niemi:2012aj,Snyder:2020rdy,Nijs:2020roc}\@. We hence see that while the influence of $\sf{}$ is moderate over the complete centrality range (left panel) the contribution becomes relatively important in ultracentral collisions (middle and right panels) because of the small absolute magnitude of $\epsilon_2$ and $\epsilon_3$ in ultracentral collisions. The subnucleonic structure, varied by varying $n_c$ is however only important for very peripheral collisions and more so for $\epsilon_3$ than for $\epsilon_2$\@. To describe the ultracentral $v_2\{2\}$ it is hence essential to estimate $\sf{}$ accurately, which in \tr{} is done by including transverse momentum fluctuations in the Bayesian analysis (see also Section \ref{sec:design} and \cite{Bernhard:2018hnz})\@. This can likely explain the difference with \cite{Carzon:2020xwp}, which used an older Bayesian analysis \cite{Bernhard:2016tnd} without transverse momentum fluctuations. Similar to the event plane angle correlations we hence see again the power of a full global analysis that links heavy ion physics across observables.

Several observables have not been measured in the ultracentral region, including the mean transverse momentum as advocated in \cite{Gardim:2019brr}\@. In Fig.~\ref{fig:ultracentral} we show the $\langle p_T \rangle$ for increasingly ultracentral collisions together with the multiplicity as well as the anisotropic flow coefficients $v_n\{2\}$\@. All these observables are statistically relatively cheap to compute, but we note that to obtain results for the 0--0.001\% centrality class it was still necessary to obtain 33M \trento{} events for each of the 20 parameter sets and subsequently sample the 0--0.01\% highest entropy events with a 100 times higher rate with respect to the other events (as stated we used 0.5M hydro events in total)\@. As expected the multiplicity increases relatively steeply towards more central events. This increase can in principle be compared with the multiplicity beyond the `knee' in the V0 amplitude in \cite{ALICE:2013hur}, which is also shown in the figure for 2.76 TeV collision energy \cite{ALICE:2013hur} (for 5.02 TeV the V0 distribution suffers from pile-up in this region and is harder to compare)\@. Note that our model in this work is boost invariant and it is hence appropriate to compare our mid-rapidity multiplicity and mid-rapidity centrality selection ($|\eta|<2.4$) with the ALICE V0 distribution (which uses the multiplicity/amplitude and centrality selection both at forward rapidity)\@. Ultracentral multiplicity has also been studied at RHIC \cite{Carzon:2021tif}, though here it is problematic that the STAR data is only available in raw form (the V0 data is also raw, but as a ratio with the 1--2\% class detector effects should be suppressed)\@. We do not find a significant change in any $v_n\{2\}$ (with $n \leq 4$) downwards of 0.5\% centrality.

The mean transverse momentum in extremely ultracentral collisions is especially interesting, since for these centrality classes the initial volume is approximately constant \cite{Gardim:2019brr}\@. If one furthermore assumes that the total energy $E$ and entropy $S$ per unit rapidity are approximately conserved then the increase in $\mpt{}$ is directly related to the increase in $e/s$, with $e$ and $s$ the initial energy and entropy densities that can potentially be interpreted as being directly related to the temperature. In reality the energy per rapidity decreases due to longitudinal work and entropy increases due to viscous corrections. Our results from the 1--2\% centrality class to the 0.1--0.2\% centrality class show an increase of the $\mpt{}$ for charged hadrons from approximately 715 to 724 MeV, which is in agreement with \cite{Gardim:2019brr}\@. We note that curiously there is a small but significant decrease in the pion $\mpt{}$ going from centrality 5\% to the 1--2\% class.

It is an interesting question how the extremely central observables are affected by a centrality selection which is different from the rapidity region where the observable is measured. This avoids auto-correlation effects from (rare) fluctuations and is typically what is done in experiment by e.g.~the ALICE forward V0 detector. To simulate this we redid our computation with a centrality measure given by the multiplicity at $|\eta|>2.0$ (we cut of the particlization at $|\eta|=2.5$ in our simulations, so we cannot use the 'true' V0 ranges) and show this in the bottom row of Fig.~\ref{fig:ultracentral}. As expected, the increase in multiplicity is now weaker, since  the multiplicity observable now decorrelates with  the centrality selection. The reduction of the ultracentral effect on the $\mpt{}$ is even stronger, which can likely be explained because the dependence of $\mpt{}$ on centrality is much weaker than for the multiplicity.
Note that our hydrodynamic model assumes longitudinal boost invariance, so any dependence on the centrality selection is due to thermal sampling during the particlization. Realistically the plasma itself also fluctuates (see e.g.~\cite{Bozek:2010vz,Ke:2016jrd}), which would lead to stronger effects from the centrality selection than in our boost invariant model. To avoid this it is likely essential for extremely ultracentral observables to use the same centrality rapidity interval as in which the measurement is made (as in the top row of Fig.~\ref{fig:ultracentral}).

\section{Bayesian analysis using OO simulated data}

In the previous section, we showed several predictions for the oxygen runs at the LHC and RHIC\@.
While this is very interesting and important, as it allows for a good test of the current model, one can answer an additional question.
Assuming, as we have been so far, that the soft sector of OO collisions can be described by the same hydrodynamical model as for PbPb, one can wonder whether the addition of OO data can improve the constraints on the parameters such as those obtained from PbPb alone.

One might hope that OO yields constraints orthogonal to the ones already given by PbPb because OO is a much smaller system, but it is not a priori obvious that this is indeed the case. In this section, we will obtain an answer to this question, by using simulated data for OO at $\sqrt{s_\text{NN}} = 7\,\text{TeV}$ generated by the model itself (this is similar to a closure test \cite{Nijs:2020ors,Nijs:2020roc,JETSCAPE:2020shq,JETSCAPE:2020mzn})\@.
We will perform Bayesian analyses for 20 different sets of simulated data, where the simulated data come from high statistics runs of the model itself using the parameters indicated in Fig.~\ref{fig:chainspbpb}\@. Since we assume consistency between OO and PbPb, the `true' value should be one of the points in the posterior of our fit to PbPb data, so it makes sense to use parameter settings which are randomly drawn samples from the posterior (see Fig.~\ref{fig:chainspbpb})\@. We have to make an additional assumption on the uncertainty of the simulated data. Estimates for the statistical and systematic uncertainties of the OO observables we would like to include in the fit have recently become available \cite{Brewer:2021kiv,Altsybeev:2751545,ALICE-PUBLIC-2021-004}\@, but here we make the simplifying assumption that the relative experimental errors are the same as for the corresponding observables in PbPb\@.
The observables we have used in the simulated data OO fits are the same as the ones used in the fit using only PbPb experimental data, except that for $p_T$-differential anisotropic flow we have only used the bins bounded by $(0.5, 0.75, 1, 1.4)\,\text{GeV}/c$, and we only fit pion $v_2\{2\}(p_T)$ (as shown in Fig.~\ref{fig:design})\@.

\begin{figure}[ht]
\includegraphics[width=0.49\textwidth]{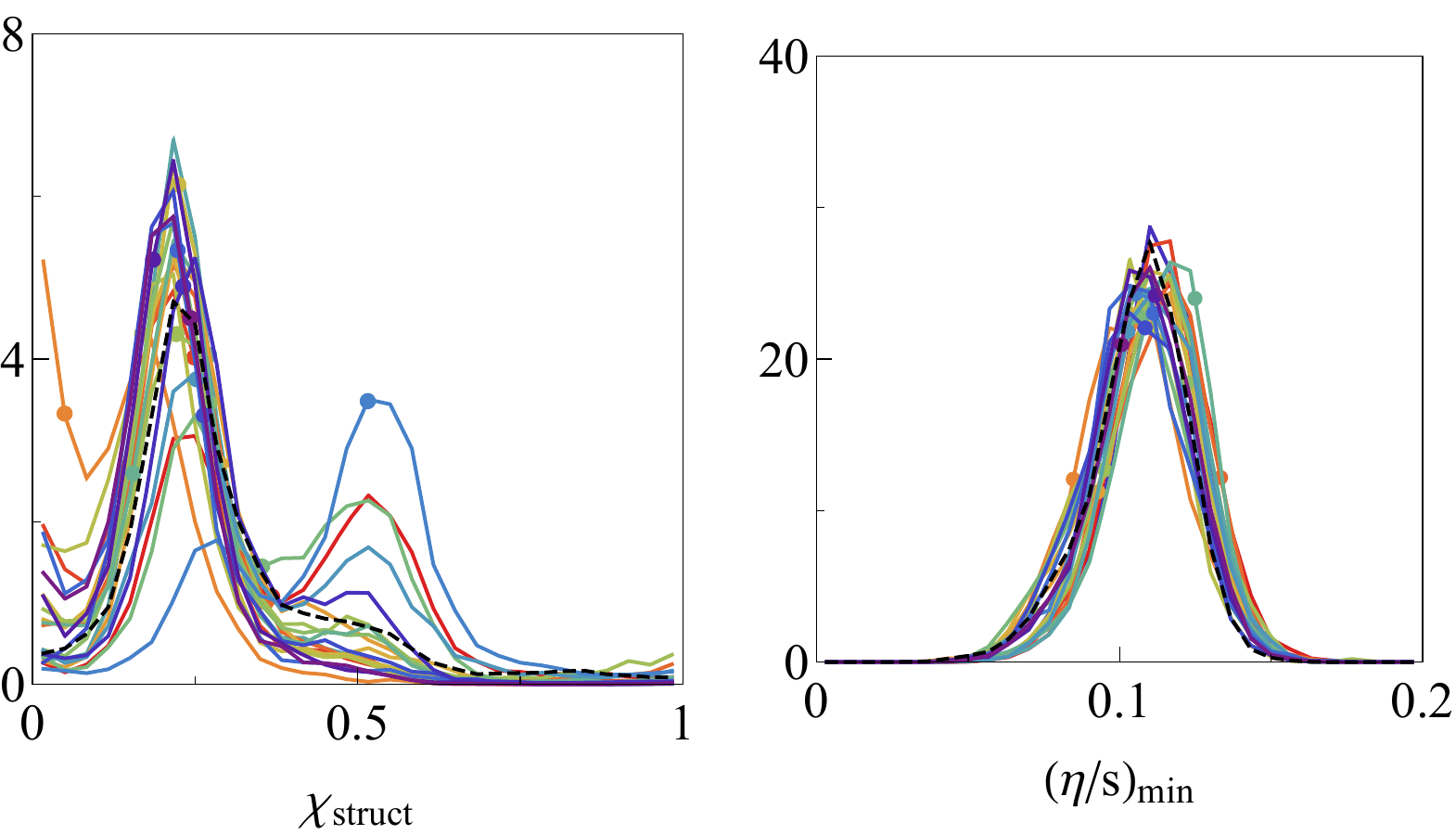}
\caption{\label{fig:chainexamples}We show some 1D posterior distributions for the 20 Bayesian analyses including OO simulated data (solid, colored) together with the posterior distributions including only PbPb data (dashed, black)\@. We show the nucleon substructure parameter $\chi_\text{struct}$ (left) and the curvature of the shear viscosity to entropy density ratio $(\eta/s)_\text{min}$ (right)\@. It can be seen that the distributions for $\chi_\text{struct}$ are all fairly close together, whereas those for $(\eta/s)_\text{min}$ show more variation. For $\chi_\text{struct}$, the distribution is narrower than the PbPb result, which can be easily seen because the peaks of the normalized distributions are taller than the PbPb result.}
\end{figure}

In Fig.~\ref{fig:chainexamples}, we compare the results of the fit which only includes the PbPb experimental data to the 20 fits which also include simulated data for OO, where in both cases we use the fits that also vary $cent_\text{norm}$\@. In the left panel, we show the posterior distribution of the $\cs{}$ parameter, whereas on the right we show $\eos{}$\@. There are two interesting observations. Firstly, comparing the PbPb fit to the simulated data fits, one can see that the simulated data fits are narrower and shifted for $\cs{}$ and virtually unchanged for $\eos{}$\@. This implies that the determination of $\cs{}$ becomes more precise with the inclusion of simulated data, indicating that under the assumptions we made, the inclusion of real OO data would also improve the precision of the fits. The second observation is that the $\cs{}$ distributions do not all have their peak in the same place, whereas the $\eos{}$ distributions are almost on top of each other. Recall that the parameters for the simulated data are drawn randomly from the PbPb posterior. This means that, as one can see in Fig.~\ref{fig:chainspbpb}, the `true' parameter values for the simulated data points are spread evenly according to the PbPb posterior distribution.
Naively, one might think that the fits to the simulated OO data should show reconstructed values centered around the `true' values, and therefore one may wonder why the simulated data posteriors for $\eos{}$ are spaced so close together. The explanation for this is that while for the simulated OO data we know the `true' parameters, this is not true for the PbPb data, which is still taken from the actual experiments. Both the PbPb data and the OO simulated data pull on the parameters, and the relative strengths of the pulls of each dataset determine the final posterior distribution. This leads us to the interpretation that the pull of OO on $\eos{}$ is small relative to that of PbPb, whereas for $\cs{}$ the pull of OO is larger.

\begin{figure*}[ht]
\includegraphics[width=0.99\textwidth]{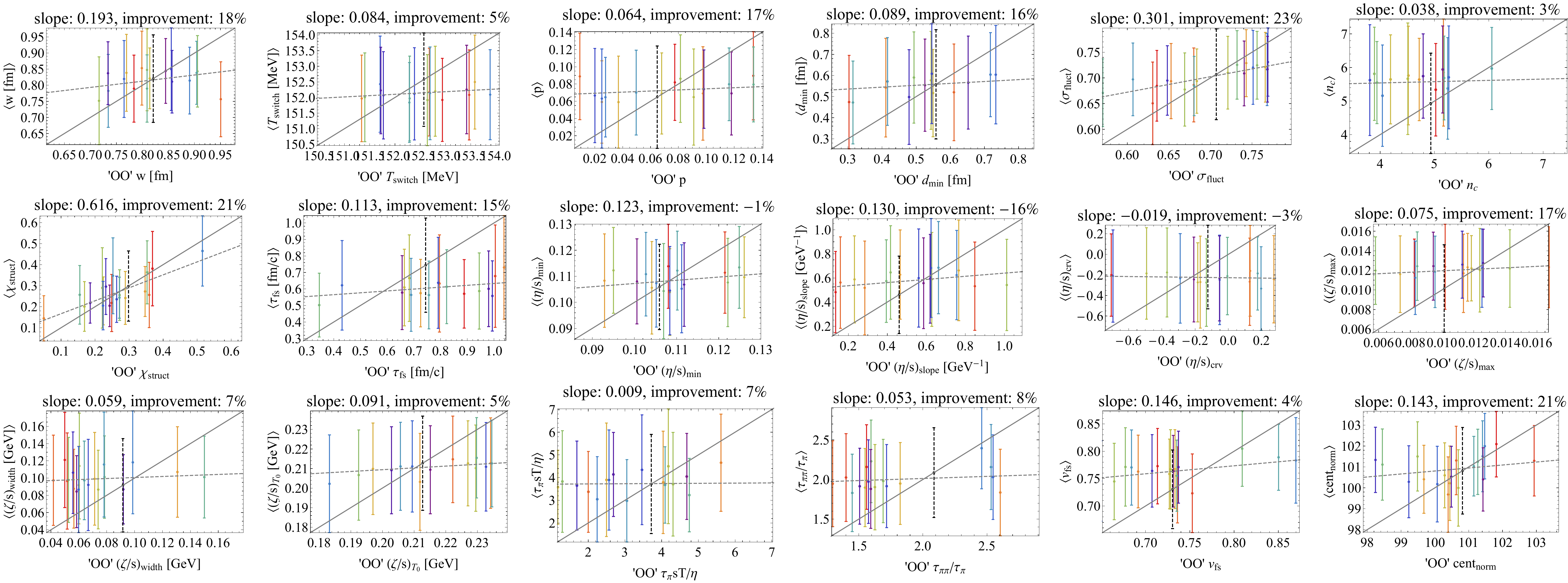}
\caption{\label{fig:chainimprovement}For each parameter in our model, we show the original PbPb fit (black), as well as the results for the 20 fits which include the OO simulated data (colored)\@. The horizontal position of the PbPb data point is given by its mean, whereas the horizontal positions of the data points which include the simulated data are given by their `true' values. We also show a linear function fitted to the simulated data points (dotted), as well as a line with slope 1 (solid)\@. The value of the slope of the linear function is also quoted, together with the improvement of the posterior uncertainty.}
\end{figure*}

Fig.~\ref{fig:chainimprovement} explores these two observations for all the parameters in our model. The black points indicate the expectation value and precision of the parameter using just the PbPb experimental data. In color, we show the various fits using the 20 OO simulated data sets as well as PbPb experimental data, where the horizontal position corresponds to the `true' parameter value that was used to generate the particular simulated OO data set. Also shown is the slope of a linear function fitted to the OO simulated data fits. If the parameter can be fully %
constrained by OO data with much higher precision than from PbPb data one would expect a slope of unity, whereas if PbPb data is more constraining one would expect a slope closer to zero. The slope hence quantifies the discussion above about the relative pull of the OO simulated data compared to the PbPb data.
In addition, we show the improvement $I$ in the size of the standard deviation of the simulated data posteriors with respect to the PbPb posterior, where we define $I$ as
\[
I = \sigma_\text{PbPb} / \langle\sigma_\text{PbPb $+$ OO}\rangle_\text{OO simulated data sets} - 1,
\]
where the average is taken over the posteriors of all OO simulated data fits, and $\sigma$ denotes the standard deviation of the posterior. As we can see, the uncertainties in most parameters decrease modestly with the inclusion of OO simulated data, as can be seen by the relatively small but mostly positive values for the improvement. This indicates that, under the assumptions mentioned in this section, the inclusion of real OO experimental data in future analyses could be beneficial to the precision of the analyses.
We can also see that the correlation varies from parameter to parameter, indicating the different relative pulls between PbPb and OO data. Let us also note that correlations between parameters are important. The parameter $d_\text{min}$ is not used in OO collisions, however its determination does benefit from the inclusion of OO simulated data. The reason for this is that it turns out that $d_\text{min}$ is strongly correlated with $\sigma_\text{fluct}$ (see Fig.~\ref{fig:design}), which is also improved by the addition of OO simulated data. Through the correlation, an improvement in one of the two parameters then translates into an improvement in the other as well.

\section{Discussion}

We performed a Bayesian analysis where we fit PbPb experimental data to a sophisticated theoretical model, where we substantially reduced the statistical and emulator errors with respect to \cite{Nijs:2020roc,Nijs:2020ors}\@. This results in reduced uncertainties in the posterior distributions for the varied parameters. By sampling 20 random parameter sets from the resulting posterior and performing high-statistics runs for each parameter set, we subsequently were able to make precise pre- and postdictions for both PbPb and OO collisions. Because the parameter sets with which these results were computed are randomly drawn, we are able to assign systematic errors to the results which arise because of the uncertainty in the parameters, whereby this also fully incorporates correlations between the parameters.

An interesting result from these high-statistics runs is a possible resolution of the ultracentral $v_2$ to $v_3$ puzzle. It turns out that to correctly describe $v_2\{2\}$ and $v_3\{2\}$ for ultracentral PbPb collisions, the magnitude of initial state per-nucleon fluctuations in entropy deposition $\sigma_\text{fluct}$ is important. As mean transverse momentum fluctuations $\delta p_T/\langle p_T\rangle$ are sensitive to $\sigma_\text{fluct}$ (see Fig.~\ref{fig:design}), including $\delta p_T/\langle p_T\rangle$ in the set of observables used to fit the parameters improves the results for ultracentral $v_2\{2\}$ and $v_3\{2\}$ as well. %

We then used the 20 randomly chosen parameter sets as simulated data to test what the effect of adding future OO data at $\sqrt{s_\text{NN}} = 7\,\text{TeV}$ would be on the precision of Bayesian analyses.
This showed that such an addition would most often yield an improvement in the precision of the extracted parameters, albeit most often a moderate improvement and unfortunately no significant improvement for first or second order hydrodynamic transport coefficients. One should however keep in mind that this is true working under the assumption that OO can be described by the same hydrodynamical model that also describes PbPb\@. This assumption can be violated by several causes related to the smaller size of the OO system, such as the need to simulate hard processes in addition to hydrodynamics, or that OO should be simulated using $3+1$D hydrodynamics instead of $2+1$D\@. The presented precision of our OO predictions, following from simply assuming the \tr{} model, will hence be an extremely interesting test. Even if it may not improve the precision of the parameters much we like to point out that the parameters are already rather precise and corroboration of the model using OO collisions will be a significant achievement which increases the trust in the precision of the model.

It was mentioned above that one reason one might expect that OO yields constraints orthogonal to the ones obtained from PbPb is that OO is a much smaller system. A valid point one might raise is why we are then comparing PbPb and OO to PbPb alone, and not PbPb, $p$Pb and OO to PbPb and $p$Pb\@. After all, $p$Pb is also a small system, so one might expect that any constraints obtained from OO can also be obtained from $p$Pb\@. In \cite{Moreland:2018gsh,Nijs:2020roc,Nijs:2020ors}, the analysis also included $p$Pb data, and there were actually extra constraints obtained by including $p$Pb in the analysis. In this work, however, with the substantial improvement in precision of the emulator, this is no longer the case, and inclusion of $p$Pb data actually makes the obtained fits less precise. Most likely the reason is the difficulty fitting the complete $p$Pb spectrum to high precision, which then pushes the MCMC algorithm to regions in parameter space where the emulator uncertainty is largest. According to Bayes theorem this is indeed the region that is then most likely (since other regions would be ruled out)\@. Nevertheless, from a modeling point of view this may not be wanted and the next necessary step would be to either improve this emulator artefact, or to improve the model itself (likely along the lines of \cite{Zhao:2020wcd})\@. Both options
would be profoundly interesting, but are left for future work. This same issue could mean that the predictions for e.g.~the OO spectra from Fig.~\ref{fig:spectraoo} are less reliable than they seem, although we stress that the OO system is significantly larger than $p$Pb and especially the centrality dependence is more robust to model than a $p$Pb collision \cite{Huss:2020dwe}\@.

Arguably the most interesting future avenue of research in the context of this work is the future measurements of the OO observables we predict. RHIC performed OO collisions at  $\sqrt{s_\text{NN}} = 0.2\,\text{TeV}$ in May 2021 and LHC has a special OO run planned in 2024 (likely at $\sqrt{s_\text{NN}} = 6.8\,\text{TeV}$)\@. It will be interesting to see how well our predictions compare to experimental data, as this will shed light on how well small systems can be described using hydrodynamics.

There are also several ways in which we can further improve our analysis. One is the lowering of our theoretical uncertainties. From Fig.~\ref{fig:emulatorerror}, it is clear that while in the present analysis most observables have statistical and emulator uncertainties small enough that the experimental uncertainty dominates, this is not true for all observables, especially for the anisotropic flow observables. Since quite a few of our parameters influence anisotropic flow (see Fig.~\ref{fig:design}), it is reasonable to assume that improved statistical and emulator uncertainty in anisotropic flow observables would result in improved posterior estimates for these parameters as well.

Another improvement that will be done in future work is a more general initial stage. The free streaming picture of the initial stage used in this work assumes zero coupling, whereas models assuming strong coupling, such as AdS/CFT, yield a qualitatively different plasma just after the initial stage. Also, while the T\raisebox{-0.5ex}{R}ENTo formula interpolates between a wide range of model behaviors, there are certain model features which it cannot describe, such as binary scaling. In future work we will attempt to address both these issues. Finally, it could be interesting to add different particlization methods, as was explored in \cite{JETSCAPE:2020shq,JETSCAPE:2020mzn}, as only including a single particlization method might bias the results.

{\bf Acknowledgements -} We thank Scott Moreland for the idea of making the number of constituents continuous. We thank Andrea Dainese, Jasmine Brewer, Hannah Elfner, Derek Everett, Umut G\"ursoy, Aleksas Mazeliauskas, Justin Mohs, Scott Moreland, Jamie Nagle, Jean-Yves Ollitrault, Krishna Rajagopal, Raimond Snellings, Mike Sas, You Zhou and especially Giuliano Giacalone for very helpful discussions. GN is supported by the U.S. Department of Energy, Office of Science, Office of Nuclear Physics under grant Contract Number DE-SC0011090.

\bibliographystyle{apsrev4-1}
\bibliography{oopaper, manual}

\end{document}